\documentclass[namedreferences]{solarphysics}
\usepackage[optionalrh,solaromanenum]{spr-sola-addons} % For Solar Physics 
\usepackage{graphicx}                    % For eps figures, newer & more powerfull
\usepackage{color}                       % For color text: \color command
\usepackage{url}                         % For breaking URLs easily trough lines
\newcommand{\aap}{    {\it Astron. Astrophys.}}

\newcommand{\apj}{    {\it Astrophys. J.}}
\newcommand{\apjl}{   {\it Astrophys. J. Lett.}}

\newcommand{\pasj}{   {\it Pub. Astron. Soc. Japan}}

\newcommand{\solphys}{{\it Solar Phys.}}

\newcommand{\ssr}{    {\it Space Sci. Rev.}}

\begin{document}

\begin{article}

\begin{opening}

\title{Reminiscences}

%%%%%%%%%%%%%%%%%%%%%%%%%%%%%%%%%%%%%%%%%%%%%%%%%%%
%% Authors Names
%
\author{\surname{Brigitte Schmieder}%$^{1}$\sep
%        I.~\surname{}$^{1}$\sep
%        I.~\surname{}$^{2}$      
       }

%%%%%%%%%%%%%%%%%%%%%%%%%%%%%%%%%%%%%%%%%%%%%%%%%%%
%% Runningheads
%
%\runningauthor{}
%\runningtitle{}

%%%%%%%%%%%%%%%%%%%%%%%%%%%%%%%%%%%%%%%%%%%%%%%%%%%
%% Affilations 
%
  \institute{$^{1}$ Observatoire de Paris, LESIA, 5 Place Janssen, 92195 Meudon, France
                     email: \url{brigitte.schmieder@obspm.fr} %\\ 
            $^{2}$ PSL Research University, CNRS Sorbonne Universit\'e, Univ. Paris 06, Univ. Paris Diderot, Sorbonne Paris Cit\'e
%                     email: \url{e.mail-c} \\
             }

%%%%%%%%%%%%%%%%%%%%%%%%%%%%%%%%%%%%%%%%%%%%%%%%%%%
%%% Abstract 
\begin{abstract}
I would like to thank Solar Physics colleagues for asking  me to write this chapter on my professional life.
%It is a very difficult task because I was not discovering a new planet or building a new instrument. 
My main interest has always been focused on the Sun, our star, from the heating of the corona, to the dynamics of prominences and their eruptions, flares and coronal mass ejections until their impact on the Earth.
I  built a new group in solar physics and gave to them my enthusiasm. They brought to me a lot of satisfaction.
We have made important advances in solar physics with a step forward to understand  the triggers of  solar activity and  their terrestrial effects.
%flares and eruptions, prominences and magnetic  emerging flux.
Our avant-garde research and discovery has opened new topics for the  solar community.
% have done many discoveries of avant-garde  before the community  address these questions.
Mixing observations obtained on the ground and in space with  theory and numerical simulations brings  a new perspective in  research.
\end{abstract}

%%%%%%%%%%%%%%%%%%%%%%%%%%%%%%%%%%%%%%%%%%%%%%%%%%%
%% Keywords
%
%\keywords{}

\end{opening}
%-------------------------------------------------
%% Figure 
%
\begin{figure}
   \centerline{\includegraphics[width=0.98\textwidth,clip=]{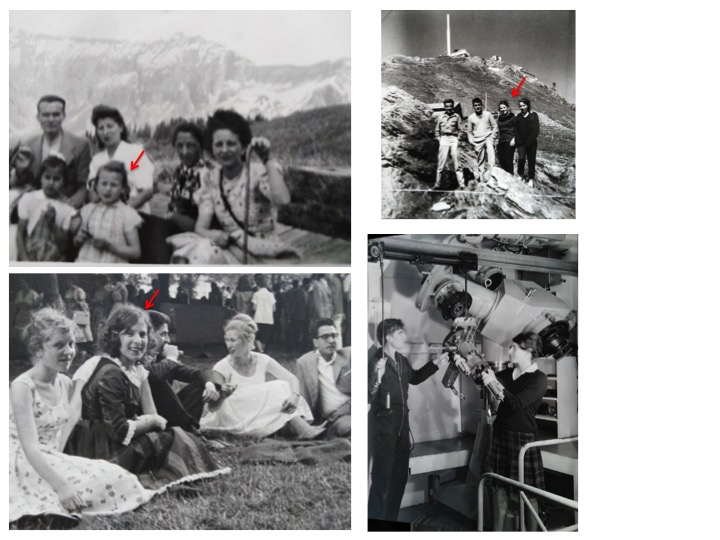}}
 \caption{Top left,  with my sister,  my father, my mother (M. and Mme. Pailhas) and two  of their friends  in Combloux (Alps), top right:  with the student  group of G.Wl\'erick at Pic du Midi, bottom left: high school  class-mates, bottom right: Saint Michel l'Observatoire, with  the electronic camera at the focus plan of the  193 cm mirror  telescope. I am indicated by red arrows.}
  \label{fig17}
 \end{figure}

%%%%%%%%%%%%%%%%%%%%%%%%%%%%%%%%%%%%%%%%%%%%%%%%%%%
%% Sections
%
% \section{}%\label{s:?} 
 \section{Early days}
 \label{s:early} 
  \subsection{Chilhood}
 \label{s:early} 

 I grew up  after the second world war in an intellectual family with my older sister. My mother  always  motivated us to learn very hard and get a good job.
 She had been a little frustrated by not being able  to do what she wanted  because of the war. She travelled before her wedding to Greece and Algeria alone. It was  each time an
 adventure for her. My father was an engineer  who   worked   in a shipping company when he was single,  landing in famous harbors in Asia (Yokohama, Hong Kong, and Saigon).  Just before the war,
 he was hired by  an oil company but was rapidly  appointed to  do  charcoal work  in  a   forest  in Burgundy. Therefore we did not go to school and 
 my mother was our teacher before we passed the  exam at  age ten    to be accepted  into  the Lyc\'ee Marie Curie in Sceaux (city located  7 km to the South of Paris).  My childhood was very peaceful (Figure \ref{fig17} top left panel). After literature and Latin, I studied mathematics and physics  with Catherine Lacombe and after the baccalaureate  I was accepted in the Lyc\'ee F\'enelon in Paris. After studying mathematics with Prof. Dixmier and mechanic of fluids with Prof. Germain at  the Sorbonne I attended the lectures of astrophysics with Profs. Evry Schatzman (the greatest physicist whom  I  ever met), Jean-Claude Pecker,  Jean Paul Zahn,  Henri van Regemorter, Roger Cayrel and G\'erard Wl\'erick.  During this time I belonged to   a group of students  and young post docs, doing sport, skiing and tennis (Figure \ref{fig17} bottom left panel). We were meeting every  Friday in a caf\'e in the Sorbonne square, ''chez Sylvain" to drink coffee and  to  decide what we would do  the next week end.  
  Ngyuen Rieu belonged to  the group and spoke to me about astrophysics. I was fascinated by  this kind of research and  by the Observatoire in Meudon  where he had a position in the   radio astronomy department.  
  Therefore when G. Wl\'erick asked if students were interested in  training, I followed him and went to  Meudon once a week.
  When I entered  his office for the first time, he told me that my name  ''Pailhas'' was familiar to him because his father, a disciple of Rodin, sculpted the bust of Mrs. Pailhas when she was very young. In fact, she was the wife of my grand uncle.
  
 After the master degree in astrophysics, I started to work at the Observatory of  Paris in Meudon (in the stables of the castle) with G. Wl\'erick  preparing  for a ''Doctorat  es Sciences''. The research for  my  thesis  effectively started  after  1968, the  year  of the  revolution of French  students when  Pierre Mein, my thesis  advisor moved from the "Institut d'Astrophysique" ( IAP) in Paris to  Meudon in the solar department.  I defended my thesis in 1977.

   \subsection{Training time}
   \label{s:formation} 
   During this long time  period with G. Wl\'erik and P. Mein, I learned astrophyiscs in many of  its different aspects. My training started by looking at the spectra obtained during the eclipse of 1961 at the "Observatoire de Haute  Provence" in Saint Michel. I discovered that G.Wl\'erick gave this task to each student  who arrived  in his laboratory.  Then, I computed theoretically the ratio of the forbidden  Ni \small{XV} lines using  atomic data  updated by Burgess and Seaton in the UK.  I wanted to compare my results with the spectral  observations of Bernard Lyot obtained at Pic du Midi.  One morning Audouin Dollfus brought me the boxes containing  the frames made by B. Lyot. One frame was missing and it was just the frame of the wavelengths of the Ni \small{XV}. We never found it. During this  time I was also working with the electronic camera invented and developed by Andr\'e Lallemand in a laboratory in Paris. We had to check the cells and had to make  a vacuum in the tube during half a day before observing. With the group of students of G. Wl\'erick we observed the planets (using UBV filters)  at Pic du Midi and at the  ''Observatoire de Haute Provence"  with the new 193 cm telescope (Figure \ref{fig17} right panels).  It was really fun to wait during  the night  for a clear sky. One winter     we built an igloo in the snow on the terrace of the Pic du Midi Observatory during our waiting periods.  I remember a few of my colleagues: Pierre Charvin, Loic Vapillon, Michel Combes, Paul Felenbok, Philippe V\'eron, and Michelle Loulergue and many young female students from the ''Ecole Normale Sup\'erieure" of  Fontenay-aux-Roses. 
   
   Back in Meudon, I had to work to build a new spectrograph in a building called  ''Petit Siderostat"  with the engineers Rolland  Hellier and Christian Coutard. It   never really  worked because 
   we put a selector of wavelengths after the grating which could not be adjustable. After  that I participated with R.Michard in the construction of the solar tower telescope   implemented by  the first  {\it Multichannel Substrative Double Pass} (MSDP) spectrograph  invented by Pierre Mein.  The observatory in Meudon developed quickly  during these years and after the construction of the solar tower we moved to new buildings adjacent to the tower.  During this time, I acquired an extensive knowledge in instrumentation, in astrophysics and in atomic physics. All these different works did not lead to any publication but they would be very helpful for my future work in science.  However, I realized only much later than  they  did not help me to
   get a  better position. I learned  at my own expense  that research not published is not valuable. It is why now I  always  push my students and collaborators  to publish their results and I am well known  to ask them  frequently    "Where is the paper?".
   
     \subsection{Heating of the corona -Thesis- 70's}
    \label{s:thesis}

    When Pierre Mein arrived in Meudon, I finally got a topic for my thesis: 
  "Solar coronal heating  by dissipation of acoustic waves: development of a radiative code and applications to observations".
   
  % No code  and no observations existed. 
  In 1948 Evry Schatzmann proposed   that acoustic wave dissipation could work for heating the corona.  Grant Athay, an American researcher confirmed this hypothesis   after theoretical computations  that such a mechanism could be efficient.  In fact nobody was really measuring the phases, velocities and amplitudes of the waves. With my observations we concluded that the waves were reflected  back in the transition region  between the chromosphere and the corona  and were mainly standing waves. They could not heat the corona. The dissipation was  too small. Before 1981  I published seven  papers  either  alone or with Nicole Mein on this topic 
     \cite{Schmieder1972,Schmieder1976,Schmieder1977,Schmieder1978,Schmieder1979,Schmieder1980,Mein1981}.
    After my thesis, Christensen Dalsgaard told me that I had to change my radiative transfer  code in LTE (local thermodynamic equilibrium) from 2  to 3 dimensions because we had discovered that the oscillations were resonant waves in the 3D sphere of the Sun. I  would have to transform my  full box of  computer punch cards with small holes to a new code.  Each day  I had to transport this box  to Orsay where the computer  ran the program overnight. Often I  came back the next day to discover that I had  forgotten  a comma somewhere. It was too tiring to  run such a  heavy code. Therefore I decided to change of topic  after my  thesis defense (see Section 2.1).
    
    About the observations  concerning the waves, it had been also a challenge to get them. The solar tower in Meudon was just built  and  the spectrograph allowed us  to get spectra of the  Mg \small{I} b line in  the low  chromosphere,  as well as  H$\alpha$ and Ca K lines in the chromosphere. After the observations, I was running a Fourier transform code and discovered that I was measuring the oscillations of the solar tower,   or of  people who were going down the stairs.  Indeed Jean Claude  Pecker  and the engineers had  an office upstairs. The interior and the external  towers were tied together in the basement by concrete.  After inspection, M. Remondet, the architect,  agreed to separate the two towers with a hammer drill.
   At  the same  time     Pierre Mein thought that the best  way to track  the granules and the chromospheric pattern was to open the slit and  use  computations to follow them. It was the birth of the MSDP.
    In fact Pierre  Mein spent one year  at Sacramento  Peak Observatory in New  Mexico and the best for me was to  go there (my first  airplane trip lasted 36 hours!). I  observed there  during the day and digitised the spectra during the night. They had great facilities that we did not have in France at that time. I met many famous scientists in  Sac Peak: Dave Rust,  Richard Altrock, Richard Canfield, and  only one woman Joan Vorpahl. When Pierre  Mein was at Sac Peak and I was   back in  Meudon, I could discuss   my problems of radiative transfer with Philippe Delache and Christian Magnan.
      
     I had to develop every thing by myself with no existing material, \textit{e.g.},  codes or observations.  Pierre Mein was there to encourage me  and gave me good advice every six months.  I would like really to thank him for his bright ideas. Pierre Mein and Nicole Mein are still working on the new generation of MSDP and very active to
    observe, enthusiastic to take part in   coordinated campaigns. I   had and still have  an intensive collaboration with them  throughout  my  research career (see my recent paper on prominence-tornadoes with them  using observations made  at the Meudon solar tower with the MSDP
    \cite{Schmieder2017a}).
    
\begin{figure}
   \centerline{\includegraphics[width=0.98\textwidth,clip=]{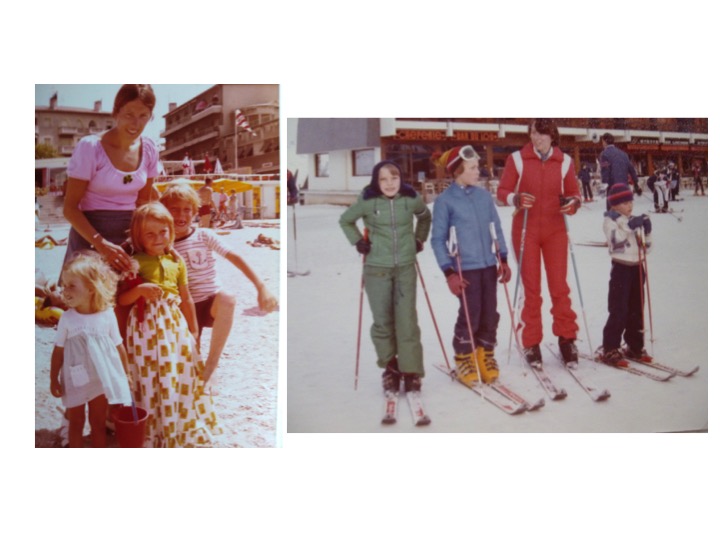}}
 \caption{Left,  on the French riviera with my 3 children: C\'eline, Anne and Laurent, right, winter sport in the Alps with Anne, Laurent and C\'eline.}
    \label{fig16}
 \end{figure}

     \subsection{Family}
 \label{s:family} 

    During my thesis I had also another task to do.  A thesis was at that time a long piece of work and could last for ten years before the change of thesis format to fit with the international PhD thesis of 3 years.      I built a family with three children: one boy Laurent and two daughters Anne and C\'eline (Figure \ref{fig16}). I  would like to thank my husband  Eric who always helped me to realize my wishes, to go to Sac Peak for one month with a baby of two  years at home. I could manage the mother's life with my work thanks  to him. After bringing  the children to   school at 8:30 am  I was back at 6 pm for the homework.

    \begin{figure}
   \centerline{\includegraphics[width=0.98\textwidth,clip=]{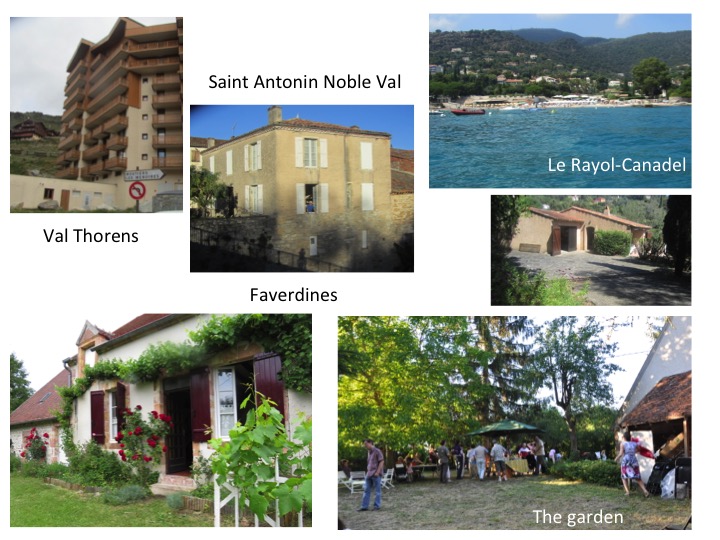}}
 \caption{Where I spend  my holidays with my children and friends and now where I am writing my papers and reviews: Val Thorens in the Alps, Saint Antonin Noble Val in Tarn et Garonne department, Le Rayol Canadel with its famous beach on the riviera, Faverdines a charming small house 
with a large garden  that many of my colleagues know.}
    \label{fig25}
 \end{figure}

  \begin{figure}
   \centerline{\includegraphics[width=0.98\textwidth,clip=]{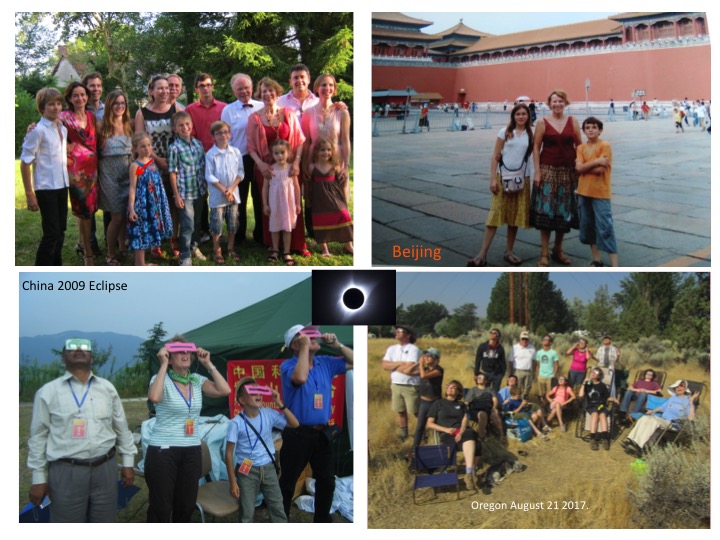}}
 \caption{Top left, my family: me,  my husband Eric   with our 3 children, the spouse  of our son and   husbands  of our  two daughters and our 8 grand children in the garden at  Faverdines,   top right:  visit of the Forbidden City with my grand children Camille and Adrien during COSPAR in 2006, bottom left, the eclipse in 2009 in  Hangshou (China) with Wahab Uddin, me, Gr\'egoire and my husband,
   %with Prof. Uddin , Gr\'egoire, Eric
bottom right: the eclipse on August 21  in  2017 in Madras (Oregon, US) 
   % in Oregon (US) 
   with Sarah Gibson and  Mark Miesh and their  family seated on chairs, behind them, standing up,  my family: Capucine, Gr\'egoire, Eric, Gautier  and me   (in red tee shirt).
}
    \label{fig4}
 \end{figure}
    My husband after his university studies, his administration school (''Ecole Nationale d'administration'' - ENA) and his military service entered in the Ministry of  health directed by  Simone Veil.
     He was hard working and  came back home only after 8 or 9 pm.  We were well organized with a nurse who received nearly all my salary. But I knew that it would  benefit me in the future.
    This situation  lasted only  during 15 years. It is a  short period of time  in a long  professional career.  I never interrupted my career for more than 6 weeks even with my three maternity leaves. Research is progressing fast and we  always need  to read and update our knowledge to be able to have the right questions.  
    We enjoyed very much to spend time with our  three children during our holidays  in our country house in Faverdines since 1966 (3 hours from Paris), in the Alps since 1980 (Val Thorens),  on the sea side in Rayol Canadel or visiting different countries (Germany, Portugal, Marocco, US)  (Figure \ref{fig25}). It was a fascinating time for us to see them growing so fast. They  get very rapidly an autonomy which helped them for their future.
    All  three  children obtained  very good jobs. Laurent,  after the ''Ecole des Arts et M\'etiers" (ENSAM), is an engineer at  the CEA (Commissariat \`a l'\'energie atomique) and is  building  the ITER facility  in Cadarache.  Anne and C\'eline  earned   very good diplomas from  the school   ''Ecole Polytechnique F\'eminine"  (EPF) and entered  in the bank research and  development department. In  industry they have to work very hard until 8 pm. They sometime regret not  having  chosen a career in the research where you have a lot of work to do but you  are your own manager. They saw their mother always working but they realized only later  the freedom that  the researchers have.
         By this time I was also involved in discussions about the ethic of  research  in a working group at the  ''Assembl\'ee Nationale"  (French Parliament) managed by Alain Lamassoure, an elected representative of  parliament.
         
  Now I have eight grand children and am happy to invite them to our different holiday resorts or  to visit   different  countries where meetings take place (Figure \ref{fig4}). They particularly appreciate watching  the solar eclipses and  like very much  the COSPAR  General Assemblies with  their space exhibitions. They discover that the English language is important to communicate with others.
  
  \begin{figure}
   \centerline{\includegraphics[width=0.98\textwidth,clip=]{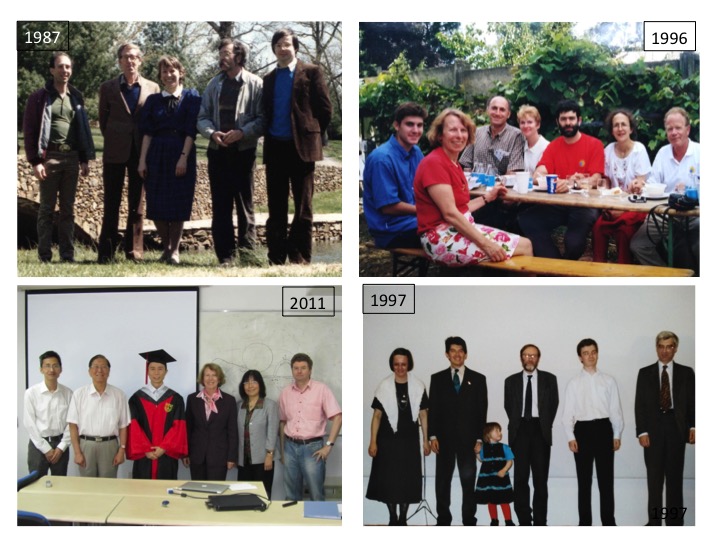}}
 \caption{Top left,
photo taken at the SMM Workshop in Airlie in 1987: Art Poland, Einar Tandberg-Hanssen, Brigitte Schmieder (BS), Guy Simon, Jean-Marie Malherbe (PhD student), top right, waiting for the eclipse in 1999 at  Szombathely in Hungary: Spiro Antiochos with his son and wife, BS, Pascal D\'emoulin with his wife, Eric Schmieder, bottom left, Guo Yang defended his thesis in Nanjing 2011 with the jury: Mingde Ding, Cheng Fang, BS, Hubert Baty,  bottom right, during the Aussois meeting in 1997, a concert was given by V\'eronique Bommier, Bernard Foing with his daughter,  Petr Heinzel, Jean-Francois Mein and Pierre Mein.
% in Tenerife after work at THEMIS: Jean-Marie Malherbe, Pierre Mein, BS, Roland Hellier.
}
    \label{fig27}
 \end{figure}
           
  \begin{figure}
   \centerline{\includegraphics[width=0.98\textwidth,clip=]{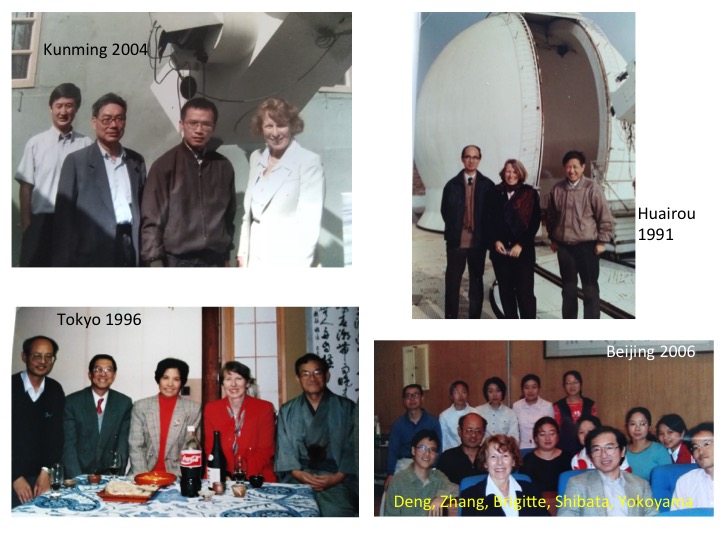}}
 \caption{Top left, Kunming in 2004: Kejun, Li, Xiaoma Gu,  Xiaoyu Zhang and Brigitte Schmieder (BS), top right, Huairou 1991: Fu Qi Jun, BS, Ai Guo Xian,  
bottom left:  dinner  by  Eijero Hiei (on the right)  in Tokyo in 1993: Hong Qi Zhang, Cheng Fang, Tang, BS,  bottom right: after  the 36th  COSPAR 2006, during  lectures  in Beijing: first row, Yuanyong Deng, BS, Shibata, Yokoyama, second row: Hong Qi Zhang, Guiping Ruan, third row: Xinming Bao and students.
}
    \label{fig10}
 \end{figure}
     
%   \centerline{\includegraphics[width=0.98\textwidth,clip=]{fig13}}\caption{top left, in Beijing 2004 Yihua Yan, BS, Ken Dere,  JingXiu Wang, top right,  2004  in Beijing BS, Sergio Dasso, Gopalswamy Nat, Jingxu Wang, Shibata, Yihua Yan, Pen Fei Chen, Guangli Huang  and Huaning Wang, E.P. Romashets,  bottom left: workshop in ISSI (Bern) in 2011: Jean Claude Vial, Duncan MaKkay, BS, Jose Ballester, Petr Heinzel, Manuel Luna, Arturo Lopez Ariste, -, Mikic,  Stano Gunar, Holy Gilbert, DeVore.  bottom right  Nanjing celebration for the thesis of Yang Guo in Nanjing  on May 25, 2011: Hubert Baty, Erdelyi, Ming Ding, Pen Fei Chen, Guo Yang, BS, student, Yu Dai, Cheng Fang.}
%    \label{fig13}
% \end{figure}

    \section{Research}
    \label{s:spatial} 
     \subsection{My concept of  research}
    \label{s:concept} 
    In 1980 the {\it Solar Maximum Mission} (SMM) was launched. The  researcher Einar Tandberg-Hanssen,   a  Norwegian  who had emigrated to the  US, and  very fond  of France, visited our laboratory  (see his memoir - \opencite{Tandberg2011}).   He was  the principal investigator  of the  UVSP  spectrometer onboard  SMM. He demonstrated to us  the interest to combine multi-wavelength observations  to understand the activity of the Sun. 
    These two events changed completely my research topics and my life.  During the preceding ten years I have  been mostly  alone in my office, not traveling, and then I discovered foreigners working together and discussing together.  
    
    It was a new world for me, more appropriate to my life concept.  E.Tandberg-Hanssen  has been the main actor to open to me the NASA  laboratories  in US (GSFC, MSFC)  and  also  to make   strong contacts in  Norway. Later on,   I was  the advisor of Jun Elin Wiik, a Norwegian student  (thesis defense  in 1993).  And finally, because of the success of Jun Elin Wiik,  I got a part time professorship in Norway from 1996 to 2006 working on filaments with Oddbjorn Engvold and his staff, \textit{e.g.},  Jun Elin Wiik and Yong Lin (PhD student). I was the second woman to  teach  physics at the University of  Oslo in Blindern.  I appreciated very much all my stays in Oslo  between the fjords and the mountains with my colleagues:  Oddbjorn Engvold, Olav  Kjeldseth-Moe  and Mats Carlson who became  a "chevalier du test vin" in Burgundy due my  sponsorship.
 
    All my research was oriented according to the collaborations  that I managed  through  bilateral research projects  which brought me to visit  many countries (US with  Einar Tandberg-Hanssen, Art Poland, Dave Rust,  and Leon Golub, Japan with Hiroki Kurokawa and Kazunari Shibata, Czech Republic  with Petr Heinzel, Greece with Kostas Alissandrakis, Georgia Tsiropoula and Kostas Tziotziou, China with Fang Cheng, Argentina with Marcos Machado and later with Cristina Mandrini, India with Venkatakrischnan and now with Ramesh Chandra in ARIES,  and recently in Korea with Tetsuya Magara at Kyung Hee University  and  Young Deuk Park in KASI.  All these researchers brought  and still bring to me a lot  of knowledge in different aspects of astrophysics. 
    
     I visited China for the first time in 1991, I was invited by Xiaoma Gu for one month  in Kunming  and again in 2004 (Figure \ref{fig10} top left panel). There I also met Jun Lin as a young student.  In China at that time  there was no xerox machine to copy the rolling paper where flare spectra of interest were recorded. I  traced the spectra on a paper and in Meudon I digitized them  by using a   machine  overlying  each curve with a stiletto and published consequently a paper on post-flare loops  \cite{Gu1992}. Before returning to France, I visited Beijing with Ai Guo Xian and Fu Qi Jun and Nanjing with   Cheng Fang  (see his memoir - \opencite{Fang2018}) who could speak  French  fluently  after his stay in France in 1987. I have a nice story about the arrival of  Cheng Fang in our laboratory.  Jean-Claude  H\'enoux, our director at that time, received a letter asking him if we  would agree to have a Chinese visitor for 2 years. We said yes. After three months he received the same letter. He replied: Yes of course. In fact one was  Cheng  Fang and the  other one  was  Feng   Cheng from Kunming. Finally   both came for two years. Later I visited China again, mainly for  giving lectures to young students (Figure \ref{fig10} right panels),  and to attend meetings, \textit{e.g}, the French-Chinese meetings in Xian and Shanghai, the 36th COSPAR General Assembly  in Beijing, the first Chinese-European meeting in Kunming in 2017.   I was co-advisor of  two Chinese PhD students: Yang Guo (Figure \ref{fig27} bottom left panel) and Jie  Zhao  who defended their theses in 2011 and 2014, respectively.
    
    I should also mention my numerous trips to India. I was first invited in  Nainital  in 2005 by Wahab  Uddin  where I met Ramesh Chandra and Navin Joshi who  are very good observers.  I got an official collaboration  (CEFIPRA) with Nainital, Udaipur, and Oooty. This allowed  me to  work and have many exchanges  with  colleagues and their students  in India. I was invited to watch the eclipse in 2009 with Gr\'egoire, my grandson  by  W. Uddin and Siraj Hasan  in Hangshou where we arrived after a long trip during the night. They  were still waiting for us. Siraj said, Brigitte did not cancel her trip, so  we have to wait for her all the night. And we were rewarded  with   clear skies  (Figure \ref{fig4} top bottom panel).
     I will refer  later  to  our  scientific results.     
    
    In addition,  we had individual  foreign visitors who also interacted  strongly with our team and our PhD students.  I will refer to  them  later to this memoir. All my PhD students  (Jean-Marie Malherbe, Pascal D\'emoulin, Guillaume Aulanier, and Etienne Pariat) could work with them.   That was an excellent training  for my students  and we maintained  good relationships and even friendships with all of the foreign visitors. Our French team developed  as an extended  group with all these researchers without frontiers.   
    With this very friendly and active ambience,  and  due to their excellent skills, all my French PhD students cited above  got a permanent job  at the Paris  Observatory and  formed the roots and core  of the   present  solar group.

%  \begin{figure}
%   \centerline{\includegraphics[width=0.98\textwidth,clip=]{fig8}}
% \caption{top left,  BS and Art Poland in Virgina automn 2006 after a SDO meeting, top  right, Buccharest  EU meeting organised by Cristiana Dumitrache: Stefaan Poedts, BS, Terry Forbes,
%bottom left: dinner in my place 2007: Tibor T\"or\"ok and his wife, Consuelo Cid, Cristina Mandrini, BS, bottom right, dinner In Beijing in October 2014   for the meeting on space  instrumentation  in China with Cheng %Fang, BS, her husband Eric, and a grand child Baptiste.
%}
%    \label{fig8}
% \end{figure}

   \subsection{Time of SMM and  J.M.Malherbe -1980's}
    \label{s:SMM}

      Jean Claude  Pecker encouraged Guy Simon  to submit    a  proposal to the Guest  Investigator  program committee of the  { \it Ultra Violet Spectrometer} (UVSP)  on  UV waves after flares.  The proposal was successful and  we got  a few weeks of observing time during several years.  Guy Simon and I  went   to the  building 7 or 21  at NASA/GSFC  in Greenbelt near  Washington DC for  planning our observations (Figure \ref{fig27} top left).  Each morning we had a meeting to decide the target of the next  day to determine the pointing of the telescope. The members of the committee hated our proposal which required to move away from the flare  site but E. Tandberg-Hanssen  regularly  reminded  them that they had accepted the proposal so it should run.  On   Sundays  Guy Simon was not allowed to enter  the NASA compound  because of his involvement in politics and I had to discuss alone the target of the day.   
       These observations were coordinated with Meudon  using the MSDP at the solar tower. Christian Coutard and Roland Hellier were our observers and they had to develop wrapping  black and white films in dark rooms every day.
 Besides it  was not easy  for Pierre and Nicole Mein to receive  the coordinates of the target. In the 1980's Internet communications were just starting. In Meudon there was only  one telephone-modem  in the computer center with which we could communicate with  the operation center at NASA/GSFC.

    The  aim of the proposal was to study the  propagation of  waves in UV   after flares. It was completely  an unknown topic  at that time, only the Moreton waves were observed.  We were really pioneers on this topic. To achieve our goal we requested that after the flare flag we moved the center of the field of view to the edge of the camera and wait   for the passage of the wave. With a Fourier transform code we tried to detect the waves. After nights and nights spent on the computer no wave was detected, due to the low cadence of the instrument. 
     However the data that we obtained were generally centered on filaments so we started to work on filaments. I met many scientists during that period with whom we  continued to work later,  \textit{e.g},  George Simnett, Art Poland (Figure \ref{fig27} top left). All became good friends  and visited France frequently. Recently I started to work again on these famous EUV  waves that  have been discovered  with the next satellite instruments in UV onboard SOHO /EIT and SDO/AIA  \cite{Delannee2014,Chandra2018}. 
           
  \begin{figure}
   \centerline{\includegraphics[width=0.98\textwidth,clip=]{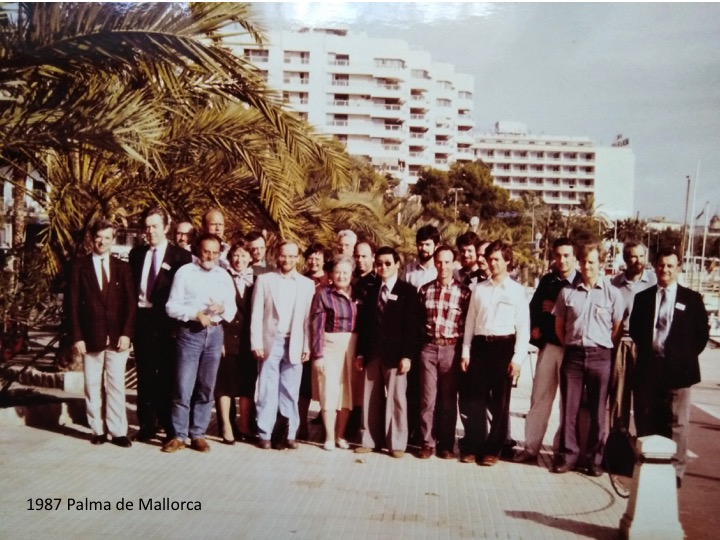}}
 \caption{Workshop  on the  {\it Dynamics and structure of solar Prominences }  in Palma  de Mallorca in November 1987,  first row,  from left to right:  Pierre Mein, Oddbjorn Engvold, Ulrich Anzer, Brigitte Schmieder, Eric Priest, Marie Jos\'e Martres, Guillermo Vizoso Miguel del Sola, Art Poland, Horst Balthazar, Alan Hood, Bogdan Rompolt, second row, Jos\'e Ballester, Eberhard Wiehr,Jean Marie  Malherbe, V\'eronique  Bommier, Jean Louis Leroy, -, Pascal D\'emoulin, -, -,  Fernando Moreno Insertis, Pierre Gouttebroze.}
    \label{fig5}
 \end{figure}

  During all these years (1980's)  I was organizing coordinated campaigns  observing  simultaneously on the ground with the MSDP either in Meudon or at Pic du Midi with Pierre and Nicole Mein.    
 With the MSDP we were observing mainly in the H$\alpha$ line  jointly  with the UVSP instruments  in the EUV lines like C IV  for  Dopplershifts of the plasma at 10$^{4}$ to 10$^{5}$ degrees.    Different topics were open to us:  the    understanding of  the dynamics of the cool  plasma in prominences, cool jets or surges, and flares. We reported our results in many conferences, \textit{e.g},  Airlie (US) in 1987 (Figure \ref{fig27} top left), in Palma de Majorca  in 1987 (Figure \ref{fig5}), in Aussois in 1997 (Figure \ref{fig27} bottom right), in Hungary
 for the eclipse in 1999 (Figure \ref{fig27} top right).

Jean-Marie Malherbe came in the group to start  a thesis in the early 1980's.   He was interested in instrumentation, physics, and computations.  He worked with us on the dynamics of prominence plasma condensations and on the dynamics of flare ribbons. We had a great chance to meet two bright  theoreticians  who visited us many times, Mike Raadu and Terry Forbes.  They brought innovative ideas during  fruitful discussions. Our measurements   from the MSDP were quantitative and not only descriptive,  therefore they were interested to check their ideas or computations.  
A very intensive collaboration with theoreticians started  at that time. It was the seed for the birth of the Meudon solar  MHD group  in the future.  With the  observations at Pic du Midi  combined with  the C \small{IV} data of the UVSP provided by  Einar Tandberg-Hanssen and Art Poland   we could measure the dynamics in fine structures of filaments, up and down motions in the feet and horizontal flows along the axis of the filament, rotation along the filament axis in case of disturbances,   and  oscillations in filaments \cite{Martres1981,Malherbe1981,Malherbe1982,Mein1982,Malherbe1983,Schmieder1984,Schmieder1985,Simon1986,Malherbe1987a}. Mike  Raadu proposed to us some simple models to explain the drainage of cool matter when the loops passed through the photosphere, the empty basket model \cite{Raadu1987a}, and also a destabilization  model of  a filament showing  twisted motions in a flux rope which lifted up as in the torus instability advanced  later on  by  Tibor T\"or\"ok  (see \opencite{Torok2011} and references therein).    The cause of the  filament destabilization  could be explained by   photospheric motions \cite{Raadu1987b,Raadu1988}.

 With George Simnett  and Einar 
 Tandberg-Hanssen  we understood that  surges  formed of   cool plasma   observed in H$\alpha$, and  jets of   hot  plasma  visible in UV and X-rays,   were co-aligned and had  similar velocities \cite{Schmieder1982,Schmieder1983,Schmieder1984,Schmieder1988a,Schmieder1988c,Schmieder1993,Fontenla1994,Schmieder1996a}. At the same time  I worked also on surges with American scientists who were visiting us: Leon Golub and Spiro Antiochos \cite{Schmieder1994a}. Both of them became good friends  of the group and we could  apply for  post doc positions  at NRL in Washington and visitorships  at  SAO at  Harvard  for our students  for their training  in MHD, \textit{e.g}, Guillaume Aulanier, Etienne Pariat, and Sophie Masson.
 
   Detection of transverse oscillations in filaments in H$\alpha$ was achieved by Bill Thompson using  MSDP observations \cite{Thompson1991}.  The SMM mission  launched in 1980 was lost during a few years before a cosmonaut going out of  the   Space Shuttle could recover it by using a  Canadian robotic  arm. It was an unbelievable rescue for   the spacecraft. Our group (Nicole Mein, Pierre Mein, Guy Simon, Jean-Marie Malherbe, and I) could   work with the data  until 1989 with our friends of  the UVSP spectrometer.  The Meudon solar tower was closed in the 1990's and  could re-open only in 2003 when THEMIS,  the French-Italian magnetograph,  on Tenerife,   the Canary islands  (opened in 1997) was  fully operating and also  after the  films recording was replaced  by J.M.Malherbe with  a CCD camera,  more convenient for the observations.
  J.M. Malherbe was  a very fast thinking student   and it is always interesting to work and discuss with him and Thierry Roudier in Toulouse  the horizontal flows below filaments \cite{Roudier2008,Schmieder2014c,Roudier2018}. 
  
\begin{figure}
   \centerline{\includegraphics[width=0.9\textwidth,clip=]{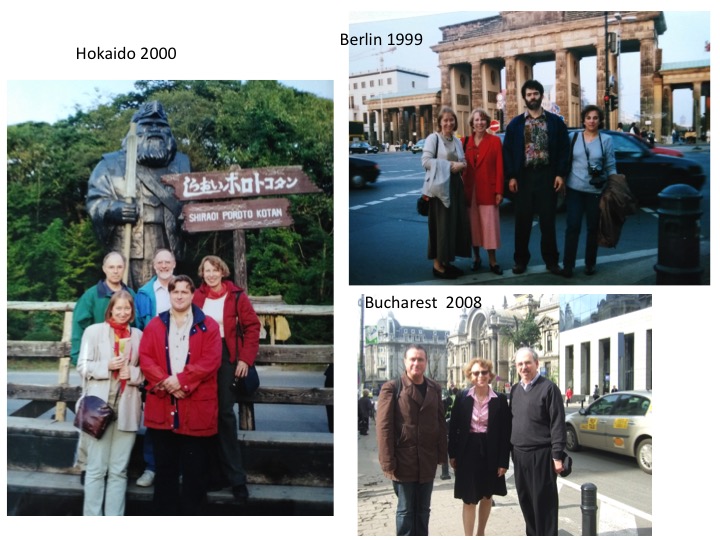}}
 \caption{Left, Hokaido in Sapporo Island visit of the Inuits after S-Ramp meeting in 2000 organized by the SCOSTEP to summarize the results of the STEP program, first row, Lidia van Driel, Stefaan Poedts, second row,  Ed Cliver, Dave Webb, Brigitte Schmieder, top right  Berlin, during  a JOSO meeting in Potsdam in 1999: Lidia van Driel, BS, Pascal D\'emoulin, Cristina Mandrini, bottom right, after the Space Weather meeting in Bucharest in 2008: Stefaan Poedts, BS, Terry Forbes.
 }
  \label{fig19}
 \end{figure}

   Our work  with Terry Forbes concerned mainly  flares. He and  Jean-Marie Malherbe   developed an  MHD model  (using the code SHASTA) to explain  reconnection in the corona and the cooling of the plasma inside the reconnected loops \cite{Forbes1986}.   The observations of the  flare ribbons  that I had obtained in Meudon followed exactly their predictions. It was the discovery of the evaporation of the chromosphere with two phases: the impulsive phase and the gentle evaporation phase in post-flare loops \cite{Schmieder1988b,Schmieder1990}.       I had the chance to meet again   Terry Forbes  in New Hampshire and during conferences, \textit{e.g},  in Romania  in   conferences organized by Cristiana Dumitrache (Figure \ref{fig19} bottom right panel).
    In the 2000's     we could confirm  the   predictions  of Forbes and Malherbe simulations concerning the chromospheric evaporation  by  using NLTE (non local thermodynamic equilibrium)  radiative transfer computations   and data of SOHO/CDS with Arek Berlicki  and Guilio  DelZanna  \cite{Berlicki2005,delZanna2006}. In the 1980's we were   pioneers  in this field of research.

   Other interesting topics involved collaborations with  our  Greek colleagues.  While working on sunspots the Evershed effect was found  to reverse in the chromosphere  compared to the  direction observed in the photosphere  \cite{Alissandrakis1988,Dere1990}. With  Georgia Tsiropoula  we  analyzed carefully the MSDP data concerning the chromospheric fine structure to understand if mottles on the disk were the same  as  spicules visible at the limb  and  had fruitful  discussions  with Petr Heinzel who was not  always in agreement with our results \cite{Tsiropoula1993,Tsiropoula1994,Tsiropoula1997,Heinzel1992}.

 \begin{figure}
  \centerline{\includegraphics[width=0.98\textwidth,clip=]{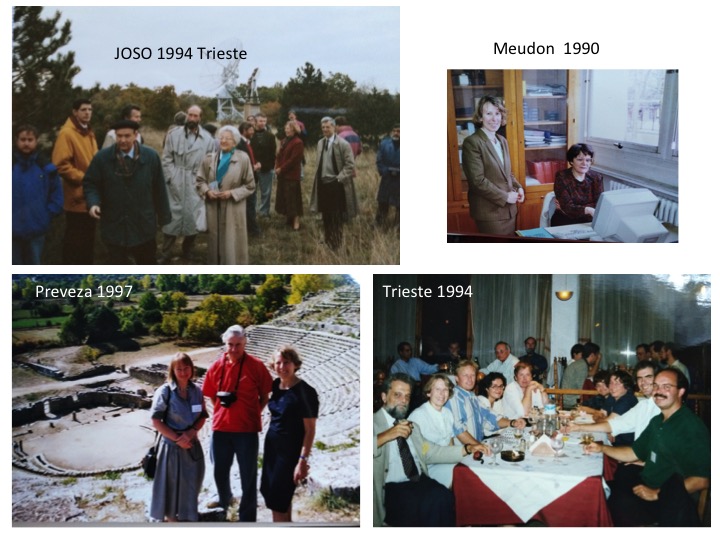}}
 \caption{JOSO group visiting  the radio telescope in Trieste in October 1994: from left to right we can recognize some scientists:  Bojan Vr\v{s}nak,  Mauro Messerotti,  Paolo Zlobec  and behind Axel Hofmann, Peter Brandt, Edith Muller -former president of JOSO-, behind, Richard Muller, Thierry Roudier and BS (president of JOSO),  Janusz Sylwester, Pierre Mein,  Lucas Vlahos , top right:  BS and Nicole Mein working in Meudon observatory,
 bottom right,  dinner during  a JOSO meeting   in Trieste in 1994: Costas Alissandrakis, BS, Paul Brekke, Greek scientist, Bob Bentley, -, Philippa Browning,  David Alexander, Bernard Fleck,  behind Meir Semel and Peter Brandt, bottom left: Lidia van Driel, George Simnett, BS  during the Preveza Euroconference in 1997 on Advances in solar Physics  visiting Dodone.
}
    \label{fig6}
 \end{figure}
        
   \begin{figure}
   \centerline{
   \includegraphics[width=0.41\textwidth,clip=]{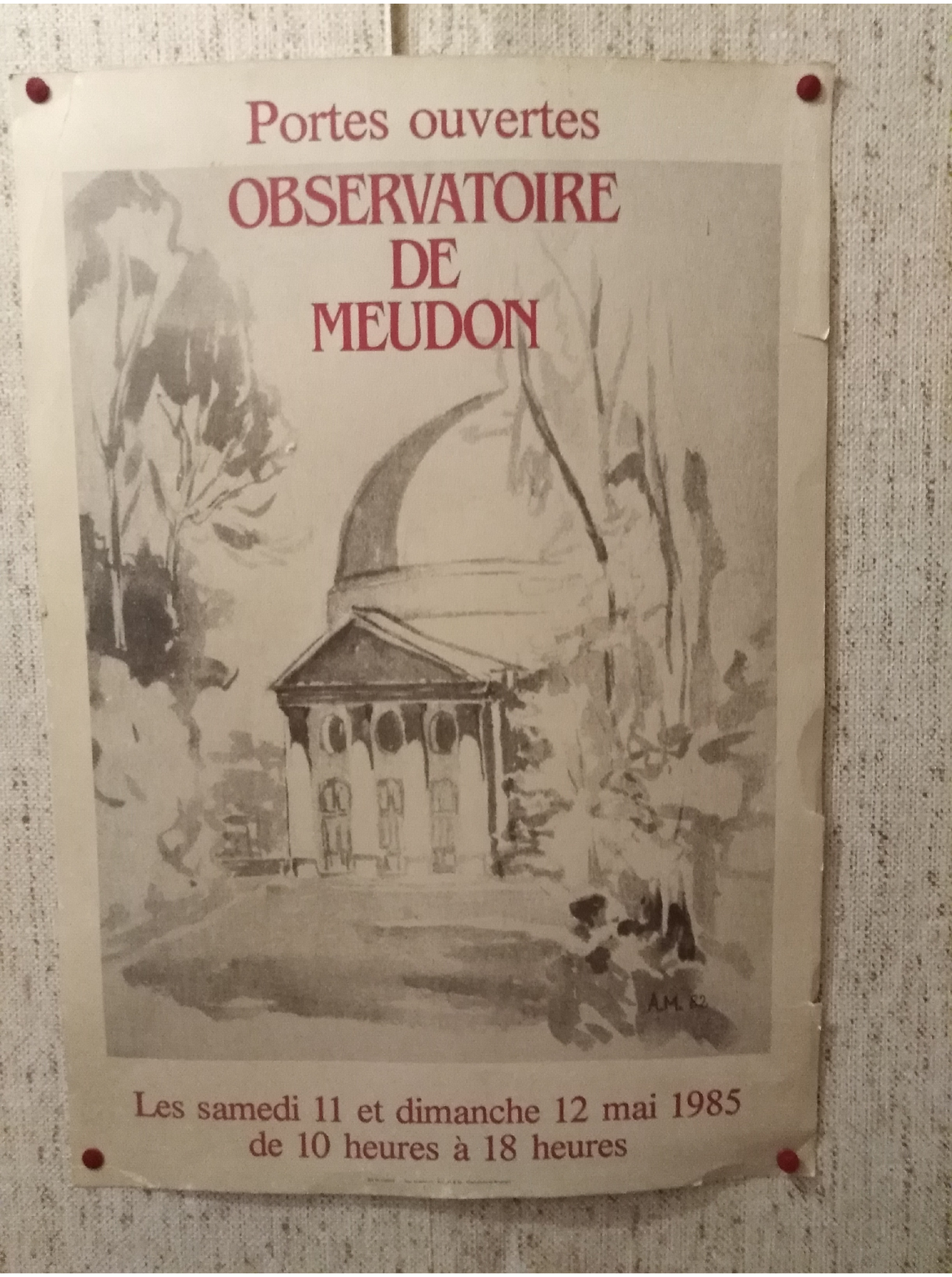}
 \includegraphics[width=0.4\textwidth,clip=]{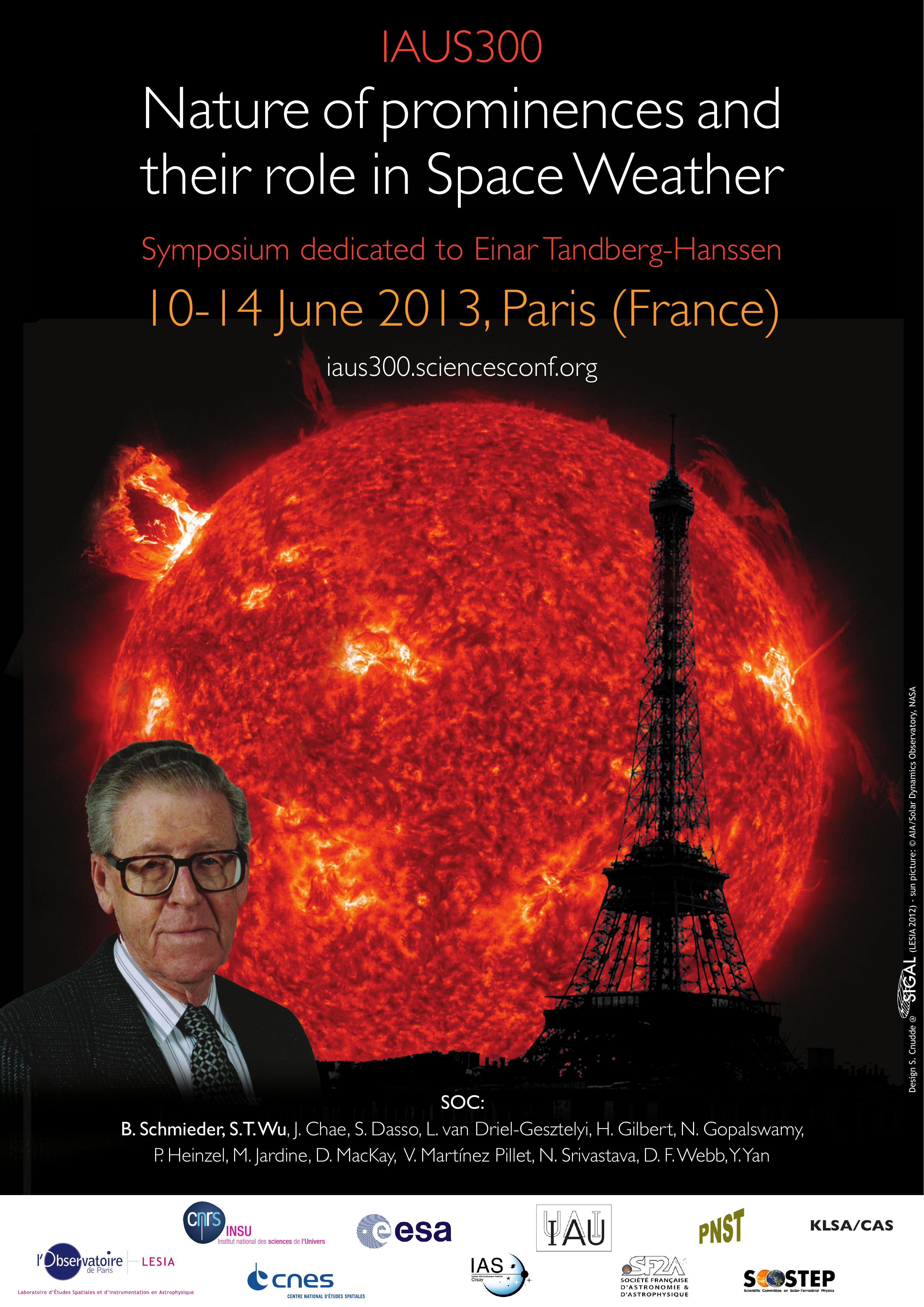}
 }
   \caption{Left, poster for the open house on 11 and 12 May 1985 in Meudon (drawn by Andr\'e Mangeney), right, poster for the IAU S300 symposium  in Paris dedicated to Einar  Tandberg-Hanssen (drawn by Sylvain Cnude).}
  \label{poster}
 \end{figure}

From 1990 to 1996, I  was elected as  vice president of the European Solar Physics Meeting (ESPM) organization.
   As ESPM vice president I organized two meetings with  president George Simnett in Catania (1993),  and in Thessaloniki (1996).  
   In 1992 I was  also elected as  president of the Joint Organization of Solar Observations  (JOSO) in Europe   until 2002.
   JOSO had been created to develop collaborations between  the solar  observatories in Europe. Therefore  I organized each year a meeting, \textit{e.g}, Trieste in 1994 (Figure \ref{fig6} top left and bottom right panels), Potsdam in 1999 (Figure \ref{fig19} top right panel). Because the attendance increased to more than 150 participants,  I wrote  proposals to European Union to be sponsored. I get two successful  EU contracts during my JOSO presidency  to organize successively  two  series of  European  conferences: three of them  on ''Advances in Solar Physics"  in Tenerife (1996), Preveza (1997), and Catania (1998)  published in PASP journals and two of them on  ''Solar cycle and Space Weather" (SOLSPA)  in Tenerife (2000) and in Vico Equenze  near Napoli  (2001) published by ESA  publications. 
   
   The main objective of JOSO was 
to involve  the solar physicists of all the European countries to create a data-base of   solar observations,  and  to define the characteristics of a future large European  telescope. Now this objective is pursued  by  the  European Association for Solar Telescopes (EAST) consortium, which has been  charged to  build  the  European Solar Telescope (EST)   (4m mirror). 
Today the plan is to  start   construction in 2021 and  to achieve the first light in 2027. Let us see what will  the future of EST be.

  Between 1980 and 1992, I was   the public outreach   manager of the Observatoire of  Paris. I was in charge  of  press releases  and the organization of  visits  to  the Observatoire de Paris in Meudon of   students
 and  the general public.
%  and the  press releases (\url{https://www.obspm.fr/les-tornades-solaires-ne-sont.html?lang=fr}). 
  I organized night shows for  journalists, lectures for  teachers on  Wednesdays and also open doors for the public  once a year. People appreciated these initiatives and  it happened that we had more than one visit of 30 persons each day. Particularly memorable   the open house  in May 1985 when 10 000 persons visited the observatory during the weekend (Figure \ref{poster}). I got funding  from the Ministry 
 of the Culture for  renovating the big dome  (over the Meudon castle)  and the refractor built by  Jules Janssen.   
 
 But after the  tornado in Paris in December 1999 the dome lost  part of its coverage and despite some ministry funding, it is still  under rehabilitation and no visits are   possible anymore.
    %with our new PhD student J.M.Malherbe
  
%  \begin{figure}
 %  \centerline{\includegraphics[width=0.98\textwidth,clip=]{fig11}}
% \caption{top left,   IAU colloquium 167 on  {\it New perspectives on solar prominences } in Aussois  in May 1997, some of them are the artists of the concert, first row : Karine Bocchialini,  Bernard Foing  (violonist) and his two children, Jean-Francois, son of P.Mein (violonist), Jean Marie Malherbe, second row : wife of Engvold, V\'eronique Bommier (piano), Dave Webb, Pierre.Mein (piano), Peter Heinzel (violonist), Iraida Kim, Zadig Mouradian, Brigitte Schmieder (B.S), third row: Oddbjorn Engvold, Vic Gaizauskas, P.D\'emoulin, Giovanni  Peres, Dave Rust, Jingxiu Wang, -,  Einar Tanberg-Hanssen, top right, dinner  in  Aussois 1997: Pierre Foukal, Dave Rust, BS, O.Engvold, his wife, E.Tandberg- Hanssen, bottom left: walking in Aussois : B.S, P.D\'emoulin, G.Aulanier and Lidia van Driel - bottom right, dinner in Aussois:Duncan MacKay, Byrne Brendon, P.D\'emoulin, Spiro Antiochos, T. Forbes, Lidia van Driel-Gesztelyi.}
%    \label{fig11}
% \end{figure}

  \begin{figure}
   \centerline{\includegraphics[width=0.98\textwidth,clip=]{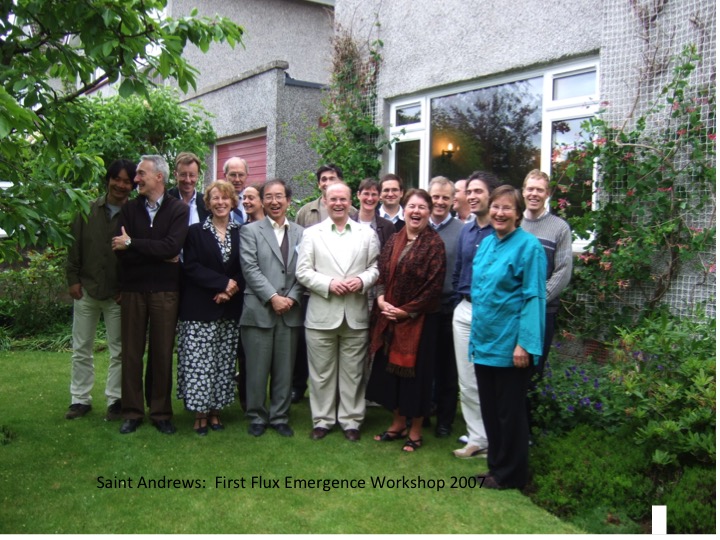}}
 \caption{St. Andrews 2007,  first Flux Emergence Workshop, in front of the house of Eric Priest, from left to right, first row: Isobe Hiroaki, Fernando Moreno Insertis, BS, Shibata Kazanari, Eric Priest, Gael the wife of Dave Rust,  Clare the wife of Eric, second row: David Hughes, Dave Rust, Laur\`ene Jouve, Evy  Kersale, Clare Parnel, Etienne Pariat, Alan Hood, Dana Longcope, Sacha Brun,  Klaus Galsgaard.
}
    \label{fig11}
 \end{figure}

         \subsection{Time of {\it Yohkoh} and  P. D\'emoulin  - 1990's}
     \label{s:colla} 
     My work  all along my career was guided by the new solar missions  launched  to resolve the problem of  coronal heating.  It is still an up-to-date topic as we heard on  TV on August 12, 2018  when the {\it Parker Solar Probe} was launched  to   approach the  Sun,  the main goals being to understand the heating of the corona, and to measure {\it in situ}  the solar wind for  future journeys of humans to  Mars or the  Moon. 
     
    After SMM,  {\it Yohkoh}, the Japanese-American mission was launched in 1991 with several instruments. I was mainly involved with the \textit{Soft X-ray Telescope} (SXT). During that time period TRACE  (1998) was dedicated to observe the Sun in UV. Both SXT and TRACE  observed partial fields-of-view and  we needed to select  the targets. I was guest investigator for both missions  and organized multi-wavelength campaigns; {\it Yohkoh} was open to foreigners only after 1993. I went to the Institute of Space and Astronomical Science in  Fuchinobe several times  learning how to process the data. It was the beginning of the solar software (SSW).
    
     I met many scientists again: Bob Bentley (Figure \ref{fig6} bottom right), Jim Lemen, Lidia van Driel (Figure \ref{fig19} left) and  by chance a few  Japanese in Mitaka:  Tadashi Hirayama,  Eijero Hiei (Figure \ref{fig10} bottom left), Takashi Sakurai, Hiroki  Kurokawa and  Kazunari Shibata (Figure \ref{fig11})  (see the paper on surges by  \opencite{Schmieder1995a}) and those on X-ray bright points  (\opencite{vanDriel1996,Mandrini1996}).  
     The main topic that we worked on was  flare-loop formation.   For the coordinated campaign with {\it Yohkoh}  we  (myself, J.M. Malherbe, and P. Mein) were observing with the MSDP at the "spectro tourelle" at the Pic du Midi  in the 1990's. We obtained very fine observations of  flare loops   with the MSDP  in June 26, 1992  to  compare with  the Yohkoh loops.    These observations  led to more than six papers with Jun Elin Wiik, Jean Marie Malherbe, Lidia  van Driel,  and Petr Heinzel  \cite{Schmieder1995b,Schmieder1996b,Wiik1996,Wiik1997,Malherbe1997,vanDriel1997}.   I  had already  started  with Petr Heinzel to work on   plasma conditions in post flare loops \cite{Heinzel1992}.  We continued and  computed the theoretical times  to cool the X-ray loops to H$\alpha$ temperature  by  radiative cooling  and conduction \cite{Schmieder1995b} and could  confirm what the model of  Terry Forbes predicted, that the hot loops visible after reconnection in X-rays  with {\it Yohkoh} were cooled down to 10$^4$ K  after a certain time and observed in H$\alpha$.
%       and  besides with  Gu Xiaoma using data from  the Kunming spectrograph.   l and I   could  confirm what the model of  Terry Forbes was predicting. A series of papers on post-flare loops appeared during this time period . 
  Other papers appeared  later on post flare  loops visible after the impulsive phase of  flares \cite{Gu1997,Schmieder1998}.
  
     However  during the loss of SMM and   before the launch of SOHO (1996), the space  data   were not easy to access  and it was the period when 
 Jean Heyvaerts, who was the  responsible professor assigning the PhD  students in different laboratories proposed to me to  be  an advisor of    PhD students.   Heyvaerts   appreciated  the way that I was training students on  solar observations  with an open view. In the  first year, I got a very bright Chinese student Z.S. She \cite{She1986} but he wanted to work only on turbulence so after one year he moved to Nice with Uriel Frish. The next year I got Pascal D\'emoulin  and  Heyvaerts told me that I should  be able to  teach  him to explain all the ''zoo of solar physics: flares, sunspots, plages, surges and eruptions''. It is with Pascal that our solar MHD group was born finally.   
For  his  PhD I had to find a theoretical group which could  give to him a proper  theoretical background. I had good relationship with theoreticians in Florence, \textit{e.g.},  Georgio Einaudi, Franca and Claudio Chuideri. Pascal was very enthusiastic to interact with them.
 However after one year no paper has emerged. My famous question "Where is the paper" was asked. At that point he  had only one  but very impressive  paper with me  and Mike Raadu on   the role of parallel and perpendicular conduction in the stability of filament  fine structures  \cite{Demoulin1987}. Because I knew that he had to publish if he wanted  a permanent position, I asked Eric Priest if Pascal could visit him in St. Andrews.  Eric Priest reported in his memoirs: "It was
a real pleasure to work with him, since he is so bright and couples a superb physical understanding
with great technical skills'' \cite{Priest2014}.  And  this is how   and  why  a solar  MHD group developed in Meudon.  Pascal and myself  were invited to many meetings, \textit{e.g.}, in  Potsdam for the JOSO meeting in 1999  (Figure \ref{fig19} top right panel) and in Hungary for the 2009  eclipse   (Figures \ref{fig27}  top right panel).

 As  a PhD student, P. D\'emoulin  began  to work with  the general topic of prominence equilibria,
the formation of dips and their support and loss of equilibrium or instability in a force-free
field \cite{Demoulin1988,Demoulin1989,Demoulin1992,Demoulin1993}.
 %Démoulin, Priest, and Anzer, 1989;
%Démoulin, Malherbe, and Priest, 1989; Démoulin, Priest, and Ferreira, 1991).
 These ideas were  further developed    by  Guillaume Aulanier in his   thesis \cite{Aulanier1998a,Aulanier1998b,Aulanier1999,Aulanier2002}. And  so  later on Pascal D\'emoulin kept  contributing  intensively to the theoretical training  of my  other PhD students.  
 By that time  (1990's)   P. D\'emoulin was also interested in the topology of the flaring active regions and pioneered research on reconnection without null points, along the  quasi-separatrix  layers (QSL) \cite{Demoulin1996,Demoulin1997}. This theory was applied to many observations, by the successive  scientists of our  group,  e.g., Guillaume Aulanier, Etienne Pariat, Kevin Dalmasse,  and also  in Argentina, China   and India \cite{Deng1999,Mandrini1996,Schmieder1997,Moore1997,Mandrini1997,Berlicki2004,Mandrini2006,Schmieder2007a}. This work continues to be developed by the Argentinian  group,   still with   linear force-free field extrapolations.
 By comparing the  similarity of the  locations  of QSLs  for a few case-studies using    linear and non-linear force-free field  approach, the robustness of 
 the QSLs,  regions of intense electric currents before flares  has been demonstrated  \cite{Mandrini2014,Chandra2011,Dalmasse2015,Joshi2019}.
  
 In the 2000's P. D\'emoulin applied the  theory of the conservation of  magnetic helicity to  active regions and showed that  the excess of  magnetic helicity in twisted flux ropes can be expelled by  coronal mass ejections  (CME). This created a stream  of various papers  in our group   (see the  Pariat  papers  in his thesis,    \opencite{Chandra2009}, \opencite{Zhao2014})  and more recently papers by  Kevin  Dalmasse \cite{Dalmasse2018}. It is still a hot topic in the group with our new PhD student Luis Linan..
          
    \begin{figure}
   \centerline{\includegraphics[width=0.98\textwidth,clip=]{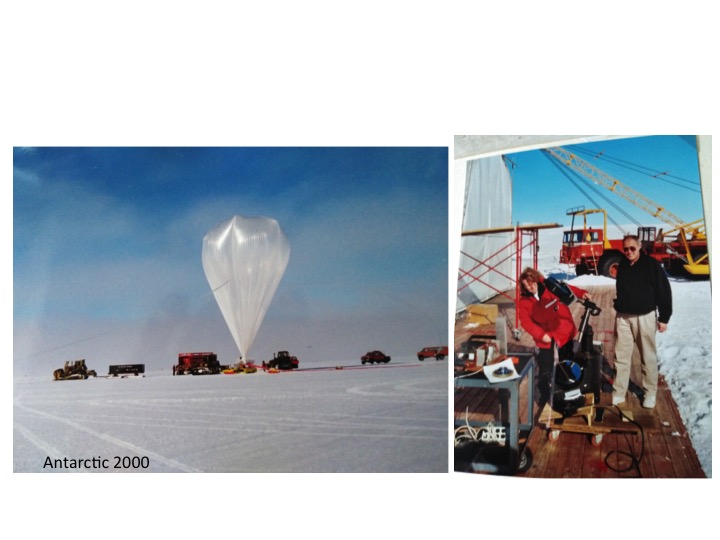}}  
 \caption{Left,  launch of the {\it Flare Genesis Experiment } hung below a balloon in MacMurdo (Antarctica )  in 2000,  right:  Brigitte Schmieder and Dave Rust adjusting the coelostat in the barn before the launch.
 }
  \label{fig21}
 \end{figure}

     \subsection{Flux emergence  - Flare Genesis Experiment-  E. Pariat - 2000's}
 The     {\it Flare Genesis Experiment}  (FGE) was a  balloon-borne telescope that made a journey of 17 days around the South Pole at 36 km altitude in the sky (Figure \ref{fig21}).  This experiment was a precursor of the  recent { SUNRISE}  instrument. FGE was a Fabry Perrot  instrument with which we could observe the chromosphere in  the H$\alpha$ line and a magnetograph which registered the Stokes parameters.  Dave Rust was the PI of the different flights of this experiment. In 1998 the balloon came  down in the French territory ''Terre Ad\'elie" in Antarctica just a few days before the  closing of the base.
     Dave Rust did not  have the official contacts between NASA and the CNES  needed for  the telescope  to be  be retrieved. Finally he phoned  me on a Sunday asking me if I could help. I called  Roger  Gendrin in Brest and everything went smoothly. The observations  were recovered and send back  to MacMurdo by using a small French airplane. In 2000 Dave  proposed to me to participate in the second flight of FGE  in case  similar problems with the French territory arose.
     I spent a month at  MacMurdo  base helping to adjust the instruments, and,  after the  launch, to define the targets for FGE, jointly with observations obtained with TRACE and {\it Yohkoh}. 
     Dave Rust was a fantastic manager and organized our stay (only six persons at the base) very smoothly. I thank him very much to give me the opportunity to go   Antarctica. 
           
 \begin{figure} 
% \centerline{\includegraphics[width=0.5\textwidth,clip=]{<fig.eps>}}
 \centerline{\includegraphics[width=0.98\textwidth,clip=]{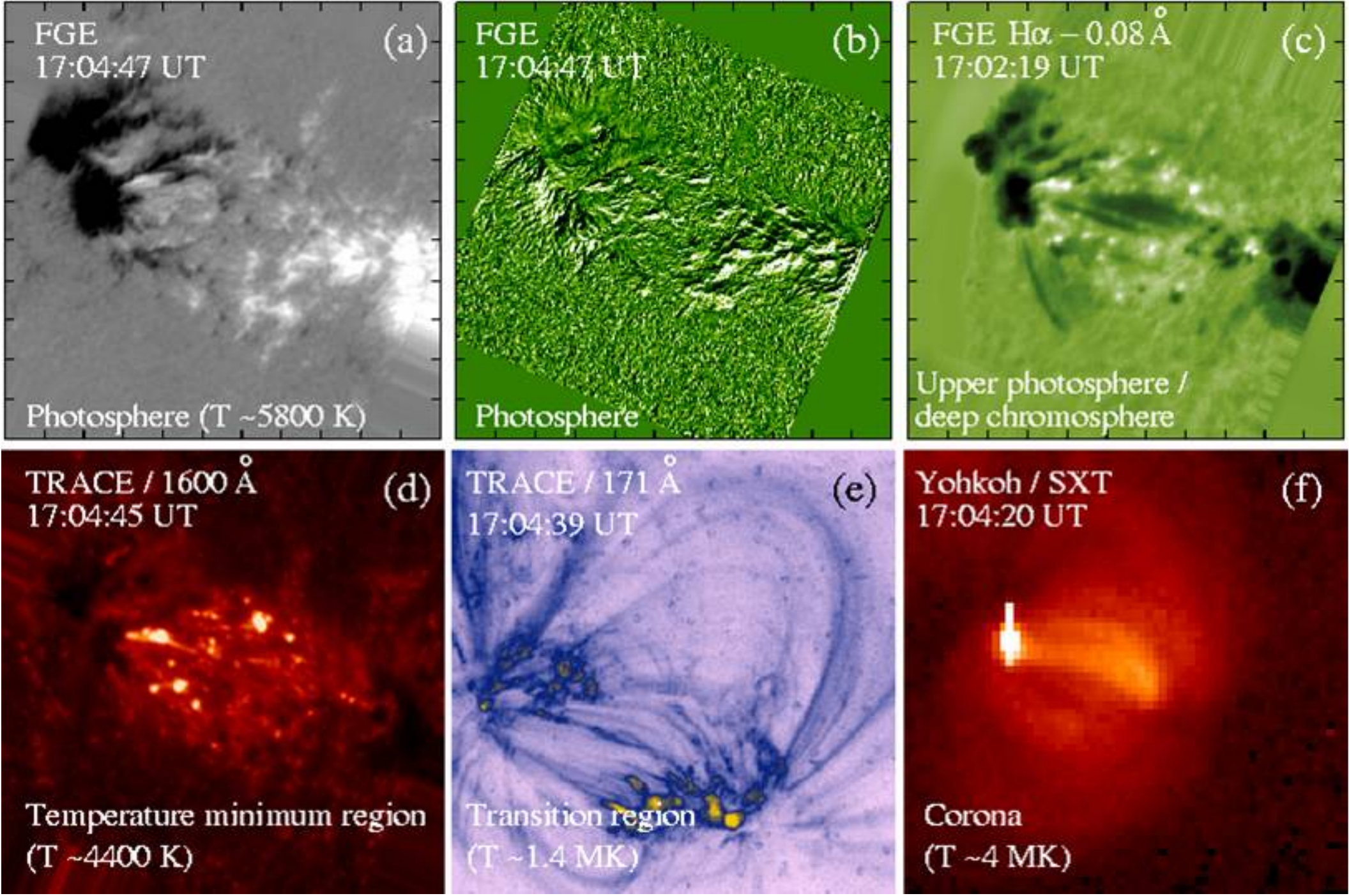}}
 \caption{{\it Flare Genesis Experiment} results: (a)  magnetic field in the photosphere,  (b)  electric currents in the photosphere,  (c) Ellerman  bombs in H$\alpha$  and  joint observations  obtained  with (d)  TRACE intensity of the transition region  (e) {\it Yohoh}/SXT in the corona, (f)  and SXT observations of very hot loops in the corona  (Pariat et al 2004,Schmieder et al 2004).}
 \label{FGE}
 \end{figure}

   We were lucky to observe newly  emerging magnetic flux during this flight.  With my new PhD student Etienne Pariat,  we could, for the first time, derive the undulatory behavior of the flux tubes as they emerged through the photosphere,   by analyzing the magnetic vectors and constructing the magnetic field in the corona using linear force-free extrapolation. We  also located the heating points where magnetic reconnection occurs and the cooling and heating loops at different temperatures over the emerging flux  region \cite{Georgoulis2002,Pariat2004,Schmieder2004a} (Figure \ref{FGE}).        
     Now with SDO/AIA and HMI  it is possible to see all the UV bursts over the full disk produced by flux emergence. My more recent  PhD students, e.g., Zhao Jie and Michalina Grubecka  continued this work  as  a part of their theses to determine where   the reconnection really occurs, in the photosphere or higher up,  based on non-linear force-free extrapolation as well as with  the  NLTE modelling of Ellerman bombs  (EBs)  provided by  Arek Berlicki and Petr Heinzel \cite{Pariat2007,Zhao2017,Grubecka2016}.\\
    We summarized the main results  of these works in a   review on emerging flux  \cite{Schmieder2014b} and  a review on UV bursts   \cite{Young2018}.  The first  workshop on flux emergence (FEW)  has been organized in St. Andrews to celebrate the success of  {\it Flare Genesis Experiment}  around Dave Rust in 2007 (Figure \ref{fig11}) and since that time a FEW is organized every two or 3 years.  In 2019 it  held in Japan  without Dave who passed away during the winter of 2019.

       \subsection{Prominence studies - Jun Elin Wiik  - Petr Heinzel  - from 1990   to now}
    \label{s:prominence} 
    In the early 1990's  Einar Tandberg-Hanssen asked me to be the advisor for  the thesis of Jun Elin Wiik, a  Norwegian student of  Eberhart Jensen in Oslo.    He recommended that we  worked on the fascinating structures called prominences which have a long history since their  discovery by  Secchi  (1875-1877 {\it Le Soleil})  and the successive classifications existing on  material protruding  over the limb \cite{Tandberg1995}.  Jun Elin   
      started to work on the characteristics  of prominence  plasma  by deriving the electron density  \cite{Wiik1992}. Then she studied the  dynamics in filaments   using observations made with the MSDP at  the "spectro tourelle"  at  Pic du Midi. We discovered  flows in both directions  along  filament threads, that would  be called "counter streaming"   flows seven  years later  \cite{Schmieder1991}.   Using the UV data of the  rocket-launched {\it High Resolution Telescope Spectrometer} (HRTS), she computed   the differential emission measure in prominences  and uncovered  their  multiple sub-resolution thread morphology  (15 to 30 threads per pixel) \cite{Wiik1993}. This  former result   has been, more than ten years later,  the basis of the development of multi-thread models \cite{Gunar2007}.       

    After her  thesis  defense in 1993  Jun Elin Wiik  got a post-doc  in Norway  and I obtained   a  part time professor fellowship in Oslo in 1996. It was the time when SOHO was just launched.  SOHO is a fantastic mission;   its coronagraph LASCO is still working in 2018.   However I was more interested  in the  spectroscopy  data obtained with  SUMER. I spent a lot of  time during my stays in Norway   processing  the data  by using the  software developed by the Norwegian group, \textit{e.g.}, Mats Carlson, Oddbjorn Engvold,  and Olav Kjeldseth-Moe.
      % During my stay in Norway I worked on prominences observed   with  the  SUMER spectrometer aboard SOHO  (launched  late 1995), and also observed with TRACE.  
     The SUMER data   were obtained during coordinated campaigns  focused on  filaments and prominences  with the Swedish solar  telescope  (SST) at La Palma,   the MSDP  at Pic du Midi, and the space instrument TRACE.
     
      At the same time  we   invited Petr Heinzel to Meudon.  After working on  prominences and  post-flare loops  with Jun Elin Wiik (see Section 2.3), he oriented his research towards  developing his  NLTE radiative-transfer  codes   to  interpret the SUMER spectra. With SUMER we could observe all the hydrogen Lyman series  lines and derive the plasma conditions in filaments, prominences and chromospheric fibrils.  Lyman lines  obtained with SUMER were a good diagnostics for  determining the  characteristics of the plasma  
\cite{Wiik1997,Wiik1999,Schmieder1999}.  
Many interesting results were obtained on eruptive prominence, multi-threads and fine threads in   prominences \cite{Schmieder2004b}. In particular we showed that   filament absorption  fine structures observed  in H$\alpha$ with  SST corresponded exactly to the darker fine threads observed  at 195 \AA\  by TRACE in the filament channel.

  \begin{figure}
%   \centerline{\includegraphics[width=0.98\textwidth,clip=]{fig9}}
   \centerline{\includegraphics[width=0.98\textwidth,clip=]{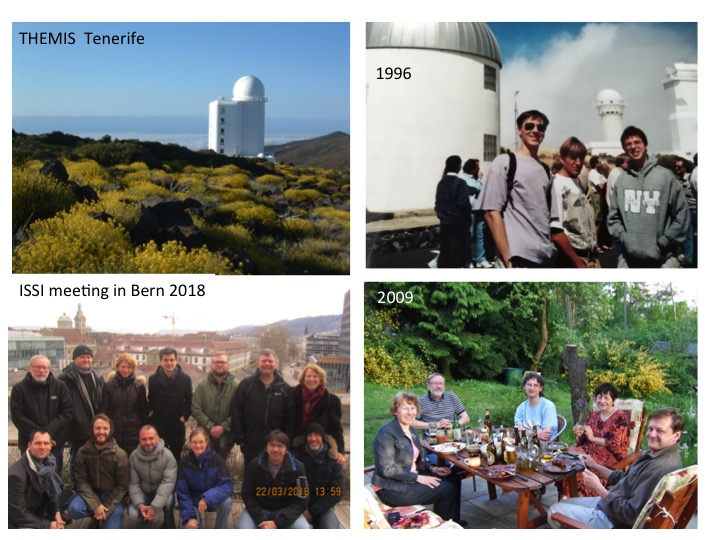}}
 \caption{Top left, THEMIS in Tenerife in 2008, top right: visit of the observatory in T\'enerife during the EU meeting in 1996: Guillaume Aulanier, Guillaume Molodij, Arturo Lopez Ariste, bottom left, ISSI meeting in Bern on {\it Paradoxes in solar physics: atypical dynamic  prominences}, first row: Stano {Gun{\'a}r}, Jack  Jenkins,  Manuel Luna, Terry Kucera, Arturo Lopez Ariste, Nicola Labrosse (PI), second row: Petr Heinzel, Maciej {Zapi{\'o}r}, Sonja Jej{\v c}i{\v c}, T.Rees-Crockford, Andrew Rodger, Duncan Mackay, BS,     bottom right: Dinner in the garden of Petr Heinzel, from left to right BS,  Petr Heinzel, Pavol Schwartz, the wife of P.Heinzel,  Arek Berlicki in 2009.}
 % the Alpen from Bern in 2011.}
    \label{fig26}
 \end{figure}

    Petr Heinzel    developed his  MALI codes to adapt them to many  different cases and structures. It was and it  still  is  a real pleasure to discuss with him. He is a hard worker and  always finds a solution.  I have always  fun to be invited for a barbecue in his place during my visit sin Ond\v{r}ejov (Figure \ref{fig26} bottom right panel).
    We published thirty  papers  on prominences  from 1998 up to now. For ten of them he  is  the first author, mainly  concerning theoretical aspects,  for ten  other papers  on observations I am the first author.  The observations are  limiting conditions of   theoretical models for prominence formation and mass loading  for coronal mass ejections during filament eruptions.  Let us quote some of them:  \cite{Schmieder1999,Heinzel2000,Heinzel2001,Schmieder2003,Heinzel2003,Schmieder2004c,Schmieder2007a,Heinzel2008}.
    The other ten  papers from the thirty  papers were led  by  scientists  of  his group in Ond\v{r}ejov, \textit{e.g.},  Pavel Schwartz, Arek Berlicki, Stano Gun{\`a}r, Jaro Dudik  \cite{Schwartz2004,Schwartz2006,Gunar2010,Berlicki2011,Parenti2012,Gunar2018}.  In particular S. Gun{\`a}r adapted  the MALI  2D-code of radiative transfer to  multi-thread structures which could be applied to the SUMER and MSDP observations \cite{Gunar2007,Gunar2008,Gunar2012}. The asymmetry in the Lyman line profiles could be explained by multi-structures  having different velocities along the line of sight.

    With {\it Hinode}, a Japanese  solar mission, high resolution  observations were obtained in  the optical range with the SOT.  They allowed  us  to see  fine structures of prominences and their high dynamic nature in Ca \small{II}  and H$\alpha$  lines. Many intriguing structures were discovered, \textit{e.g.},  rising bubbles, plumes, vertical threads.   I started to discuss with my MHD group how  we could reconcile the  MHD model of horizontal field lines with dips  developed in the Aulanier's thesis  \cite{Aulanier1998} with such observations. We invited  Jaro Dudik to model these structures and we concluded that the quasi-vertical threads were just pile-up of dips in  horizontal magnetic field lines and the bubble was a magnetized volume surrounded by a  separatrix  and a null point \cite{Dudik2008,Dudik2012}.   We demonstrated that the bubbles were not hotter than the surrounding corona \cite{Gunar2014}. Reconnection at the null point could initiate the dynamics of the plasma and  the  direction change  of the dips to form a plume. 
           
  \begin{figure}
   \centerline{\includegraphics[width=0.98\textwidth,clip=]{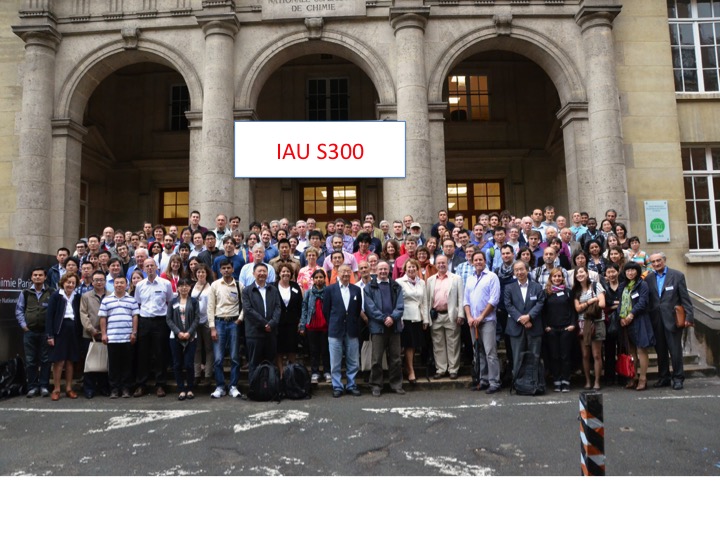}}  
     \caption{IAU S300 Symposium {\it  on the Nature of prominences and their role in Space Weather} in Paris in June 2013 in memoriam of Einar Tandberg-Hanssen: the participants in "Ecole de Chimie"  in Paris.}
       \label{fig15}
 \end{figure}

\begin{figure}
   \centerline{\includegraphics[width=0.9\textwidth,clip=]{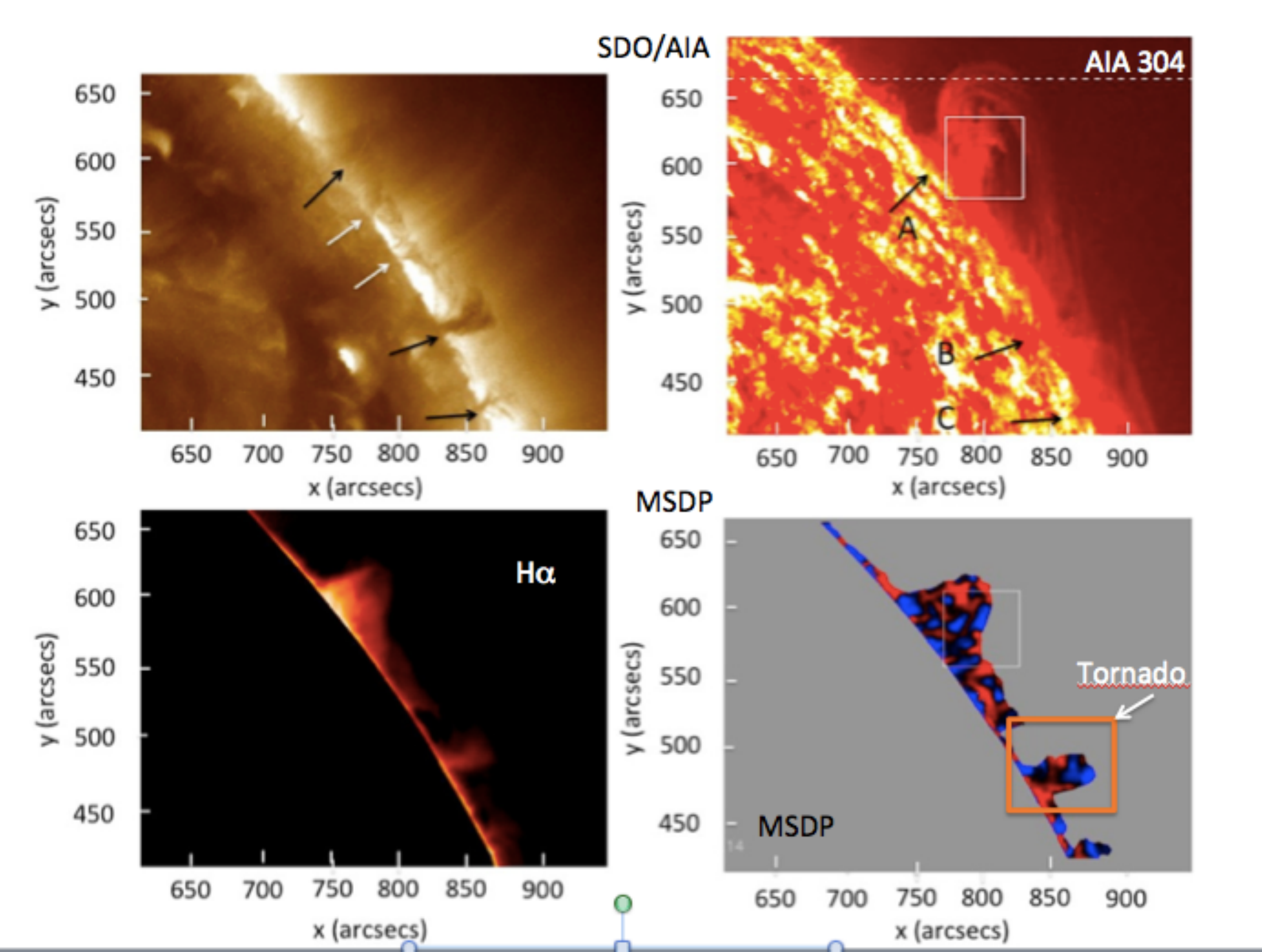}}
 \caption{Prominence observed during  a campaign of observations with IRIS, top left with SDO/AIA 193 \AA, top right AIA 304 \AA, bottom left in H$\alpha$ with the Meudon solar tower,  bottom right the Doppler shift   in the prominence and in the tornado (Schmieder et al 2014a, Schmieder et al 2017a).}
    \label{msdp}
 \end{figure}

    With the new UV  spectrograph, {\it
Interface Region Imaging Spectrograph }(IRIS),  launched  by NASA in 2013,  we continued to organize  multi-wavelength observations  during coordinated campaign with  THEMIS in Tenerife and the MSDP at the Meudon solar tower  in the years  2013 - 2016 (Figure \ref{msdp}).   
    I was observing with Arturo Lopez Ariste, then resident astronomer at THEMIS, our French magnetograph in Canary Islands  (Figure \ref{fig26} top panels). It was a great experience for all  of our  team  as they joined us successively (Nicolas Labrosse, Peter Levens (PhD student of Nicolas), Stano {Gun{\'a}r}).
    We observed more than 138 prominences  and we   confirmed the old results of  V\'eronique Bommier, Sylvie Sahal,  and  Jean Louis Leroy that the magnetic field in prominences is mainly horizontal \cite{Ariste2014}.
    
     The high spatial  resolution of IRIS  allowed us to measure Dopplershifts. Even in quiescent prominences  flows could reach  60 to 70 km s$^{-1}$ \cite{Schmieder2013a,Schmieder2014a}.
    The thesis of Peter Levens  in Glasgow was based on  these  campaigns of observation  and focused on the existence of "solar tornadoes"  \cite{Levens2016b,Levens2016a,Levens2017}.
    % the work of  Guiping Ruan,    from Shandung University on campaigns in 2017. 
    
   % Specific questions arise when looking at  SDO/AIA movies.
   %   Pierre and Nicole Mein worked on the MSDP data   (Figure \ref{msdp}. Recently 
  %  on the data  of IRIS joined with the data of the MSDP,  and  THEMIS during coordinated campaigns 
    %    I organized many observation campaigns with THEMIS since 2005    and recently specifically on prominences since 2013.  During the last period 
 Many  filaments   approaching the limb looked like tornadoes when observed in SDO/AIA  movies. They  appeared  to  rotate around their axes  like tornadoes on  Earth. Therefore they were named tornadoes.
  % Dudik2008,Dudik2012}(Ruan et al 2018). 
  However,  we  found their rotation to be  very suspicious and  that these tornadoes would be better  interpreted as the  legs of prominences observed from  a certain perspective as  they cross the limb  as  demonstrated by a 3D reconstruction of the magnetic field lines  for a  helical prominence \cite{Schmieder2017a,Schmieder2017b}. This topic  led to   press releases at the Observatoire de Paris (\footnote{https://www.obspm.fr/les-tornades-solaires-ne-sont.html?lang=fr}), at IRAP, and at the University of Glasgow in April 2018.  In these papers  Arturo Lopez Ariste processed the data of THEMIS and  concluded that the magnetic field  was  not vertical as it looks to be  in tornadoes.  
  %MHD models with dips are very appropriated to describe prominences \cite{Aulanier1998,Dudik2008,Dudik2012}.
    We continued our collaboration  with Petr Heinzel working on the Mg  \footnotesize{II} \normalsize  line profiles  observed with IRIS \cite{Heinzel2015}.  We obtained  new data  in 2017 with the MSDP at  the solar tower  jointly with  IRIS spectra. These lines give  strong constraints on  radiative-transfer models in 1D and 2D \cite{Ruan2018}.
    
       J.M. Malherbe, S.T. Wu, and myself,  organized  in Paris an IAU symposium on 'The nature of prominences and their role in space weather'', IAU S300, in June 2013, dedicated to Einar Tanberg-Hanssen who died in 2011 (Figure \ref{poster} and \ref{fig15}). Together we published a book  \cite{IAU2014}.  After intensive Team  Meetings  in Bern, at ISSI,   we wrote two  important reviews on prominences  \cite{Labrosse2010,MacKay2010}. More recently we got a new ISSI Team  project on tornadoes (Figure \ref{fig26} bottom left panel).  We are currently  writing a review on tornadoes which should be published soon (Labrosse et al 2019, in preparation).
             
  \begin{figure}
   \centerline{\includegraphics[width=0.9\textwidth,clip=]{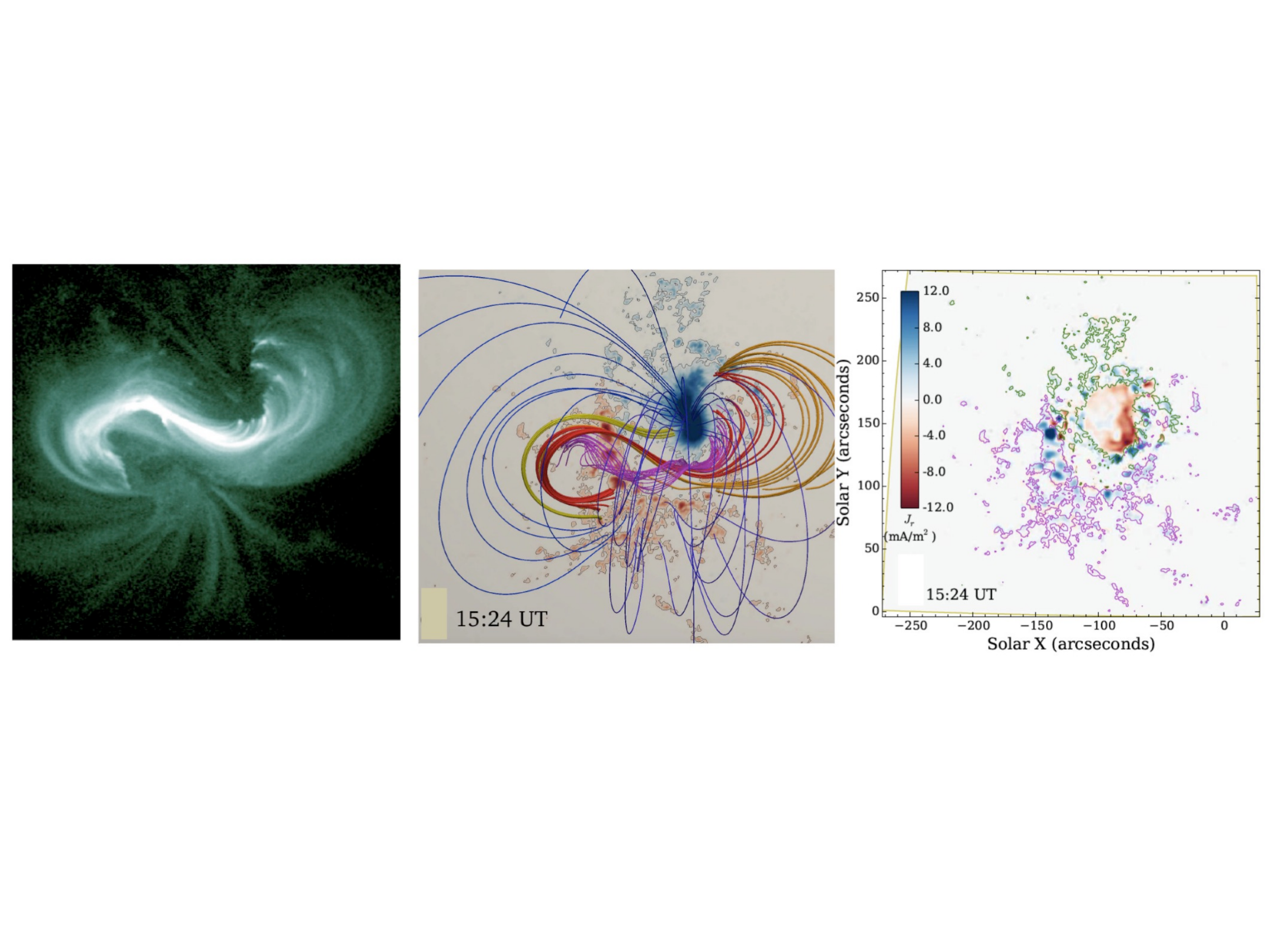}}
 \caption{Left, Eruptive sigmoidal loop in the corona observed by SDO/AIA at 94 \AA,   middle: reconstruction of the magnetic loop by a non-linear force free extrapolation, right: electric  current measured in the photosphere using SDO/HMI data (Zhao et al 2016). }
    \label{zhao}
 \end{figure}

    \subsection{Flares and CMEs   - THEMIS - 2000 - 2020}
    
    Since SMM we observed flares during  all our multi-wavelength campaigns with the MSDP,  either at the  Meudon solar tower (1980's), or at  Pic du Midi (1990's), and finally  with THEMIS after 2000 (Figure \ref{fig26} top left). I spent nearly two weeks per year on  these campaigns with Pierre and Nicole Mein.  Always a student accompanied us, \textit{e.g.},  Jean-Marie Malherbe, Pascal D\'emoulin, Guillaume Aulanier, Yang Guo, Petr Levens. It was very exiting to observe at THEMIS on the top of the mountain (Figure \ref{fig26} top panels). We were accommodated at the summit  and benefited from  beautiful sunsets.  THEMIS, designed by Jean Rayrole  with his two  post-focal instruments, the  MTR and the MSDP,   was built well after the first  designs were made  and finally their promotors died or  retired before  it  was fully operational. 
 Therefore the young generation never completely invested  its effort in the instrument  and many of them left the field. 
 %Now we have no more people working on instrumentation. 
 The  present director of THEMIS, Bernard Gelly,  is still convinced of the necessity of THEMIS which is  presently the only  magnetograph working on the ground, \textit{i.e.}  for prominences. Gelly is    working on  the installation of  adaptive optics for THEMIS which  reopens in 2019. We hope that THEMIS will  soon get  a second life.
    
    Coming back to the end of the 1990's, I had  the  pleasure to have Guillaume Aulanier as a PhD student.  
    For his thesis  (1998) G. Aulanier developed a linear force-free extrapolation code  for filament support. 
  Using his code we  showed that the differential shear above the inversion line  in an active region   induced the formation of a filament low down and that the magnetic field progressively becomes  more potential with increasing  altitude in the corona \cite{Schmieder1996c}.  By the way the importance of  3D magnetic configurations   to support prominences and to understand   their destabilization were  fully  demonstrated \cite{Schmieder1999,Kucera1999,Aulanier2002,Dudik2008,Dudik2012}.

    With the observations obtained with {\it Yohkoh}  and the MSDP at Pic du Midi,  G. Aulanier and I   worked on   flare ribbons and  flare loops to understand how reconnection  in flares can occur \cite{Schmieder1997}. Later   joint THEMIS/MSDP observations   with SDO allowed us to   investigate  the physical  conditions of  the flaring active regions  including  their magnetic topology \cite{Aulanier2012,Dalmasse2015,Joshi2016,Joshi2019,Zhao2016} (Figure \ref{zhao}).
    
         %He was a very bright student with many ideas. 
            In flares the role of bald patch  \cite{Schmieder1997,Aulanier1998c,Lopez2006},    null point \cite{Li2006,Schmieder2007b},   magnetic twist in flux rope  \cite{vanDriel2000,Canou2009,Torok2009,Guo2010,Schmieder2013a},  emerging flux \cite{Tang2000,Pariat2004,Schmieder2006,Chandra2009}, slipping reconnection \cite{Dudik2016}, flux rope reconnection \cite{Torok2011}  was  intensively analysed  in the group.   
             
      After working on so many case-studies,  G. Aulanier wrote his own MHD simulation  code ({\it the Observationally driven High-order scheme Magnetohydrodynamic code}  - OHM),  which allowed him to  develop a   3D  extension   of the  flare standard model  \cite{Aulanier2005,Aulanier2010}.  He  found  in his 3D models  innovative  solutions to explain different  signatures of the flares, \textit{e.g.}, flare ribbons, post flare loops, vortex  \cite{Aulanier2012,Aulanier2013,Janvier2014,Zhao2016,Zuccarello2017,Dudik2017} and the causes of flares-CMEs \cite{Schmieder2012}. We approached the topic of electric currents, observations and theory  by using OHM to see if there is a net current before flares \cite{Schmieder2018a}.

    The  intuitive  sense in physics and fast thinking of G. Aulanier  are  very important  in the group.  He is still interested in all  observations  made on the ground  with THEMIS,  as well as in space (from {\it Yohkoh} to SDO).  His   high  level of knowledge  leads him to find  unexpected solutions to explain  the  causes of  flares, bright points and eruptions. Therefore he is involved in many collaborations. 
    
    Up to now   P. D'emoulin, G. Aulanier and me make a very good team to lead   PhD students and  post docs in  Meudon. I can count  nearly 50  papers where  we are associated during  20 years.  Different observational and theoretical aspects of filaments and flares have been  treated.  I can quote  the PhD students that
    we supervised, \textit{ e.g.},  Arek Berlicki, Etienne Pariat, Yang Guo,  Kelvin Dalmasse on the magnetic topology of flaring active regions.  We had  and still have  today intensive and very productive collaborations with foreign countries, \textit{e.g.}, China (Cheng Fang C., Y. Tang, Li Hui,Yang Guo, Guiping Ruan), India (Gosain Sanjay, Chandra Ramesh, Navin),  Argentina (Cristina Mandrini),  UK (Lidia van Driel), and  the Czech   Republic  (Petr Heinzel, Jaro Dudik, Stano Gunar, Maciej  Zapior)   and more recently with individual  post docs, \textit{e.g.}, Stuart Gilchrist, Jie  Zhao, Miho Janvier and Francesco Zuccarello.

 Our  ideas on   physical  mechanisms for flares,  eruptions, and CMEs are  summarized in three  reviews \cite{Schmieder2013c,Janvier2015,Schmieder2015}. 
 
  %These ideas  can be put in a larger context.
 
    %During the last decade we understood that beyond the corona there is a tiny plasma of weak density called the solar wind. In the solar wind accelerated particles and  coronal mass ejections are traveling through the heliosphere.  What are the effects of all these ionized and magnetized plasma arriving in the Earth environment? That is the new question of the XXI century to resolve called  '' Space Weather''.

     \section{Space Weather}
     \label{s:SW}

    Beyond the corona  in the heliosphere the solar wind is blowing with a velocity of more than 400 km s$^{-1}$  where  coronal mass ejections and accelerated particles are traveling toward the Earth. The new problem which arises at the beginning of the 21st century is the   analysis of  their effects on  Earth. We have identified  aurora  borealis,  disruptions to the electric power grid,  disturbances in telecommunications. We should be able   to predict them  in case of extreme events like those observed on other stars.
    
    My first major steps towards 
   % I was already concerned by 
   Space Weather  research  were to organize as JOSO president  the two  SOLSPA conferences  in 2000 and 2001 (see Section 2.2)  and  to be  vice president of SCOSTEP from 2007 to 2011. I became vice president of SCOSTEP after serving as IAU representative at the SCOSTEP bureau during ten years (1996-2006). I participated in the organization of several General Assemblies,  \textit{e.g.}, Longmont, Berlin, Melbourne and in meetings,  like in  Hokaido  during the  S-Ramp meeting for the end of  the STEP program  where I met Ed Cliver and Dave Webb,  specialists of extreme events (Figure \ref{fig19} left panel).

       In 2004 the SCOSTEP bureau  asked me to form a group  with Bob Vincent to define the future program of   SCOSTEP and we proposed the {\it Climate and Space Weather of the Sun Earth System}  (CAWSES) program which was running for eight years  (from 2008 to 2016). I discovered that in France we were really  pioneers to have such program on the national level , {\it i.e.} the PNST ("{\it Programme  National Soleil Terre}'').   I get an award for  my services as  SCOSTEP vice president in June 2015.
      
       Working on the organization of the science I was also interested to see by myself  the causes of magnetospheric disturbances produced by interplanetary  coronal mass ejections (ICMEs). I worked in two groups:  a French group gathering scientists  of different communities and    in  Bern (ISSI Teams)   a group with Belgian scientists,  e.g., Luciano Rodriguez,  Stefaan Poedts,   Argentinian scientists,  e.g., Cristina Mandrini,  Sergio Dasso, Hebe Cremades and  with the Spanish group of Consuelo Cid. This work was very interesting. It showed to me how difficult is to  find  the causes of  magnetospheric disturbances. With different techniques there is still an uncertainty of more than ten hours to  be able to predict the arrival of  ICMEs  to the Earth
     \cite{Dasso2009,Rodriguez2009,Cid2012,Cremades2015,Bocchialini2018}.   
      
  \begin{figure}
   \centerline{\includegraphics[width=0.9\textwidth,clip=]{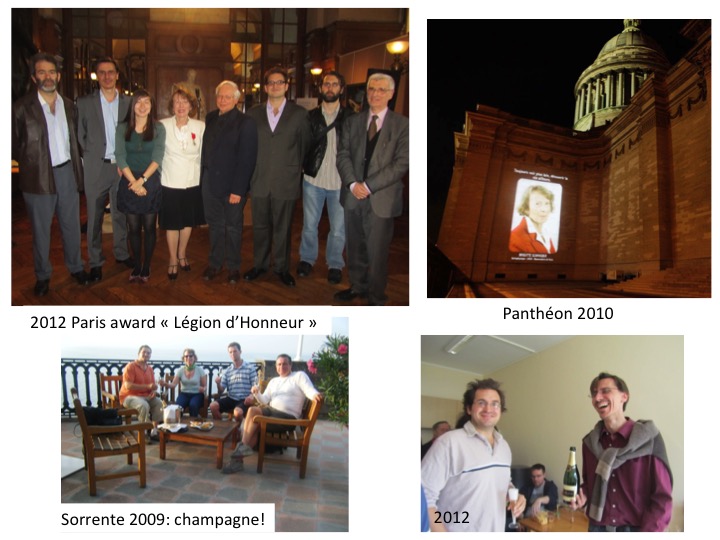}}
 \caption{top left,   Award to Brigitte Schmieder (BS) ({\it Officier de la L\'egion d'Honneur})  on June 13 2012 at the Observatoire de Paris given  by Jean Claude  Pecker. Our solar  team: Pascal D\'emoulin, Guillaume Aulanier, Miho Janvier, BS, Leon Golub, Etienne Pariat, Kevin Dalmasse, Pierre Mein, top right: BS picture projected on the facade of  the Pantheon (1000 best researchers in France),
bottom left: ESPM meeting in  Sorrento 2009, champagne for the CNRS position of Etienne  Pariat at the Observatory of Paris, BS, Daniel Muller,  Stefaan Poedts, bottom right:  champagne again in Meudon with my two former PhD students: Etienne Pariat and Guillaume Aulanier for his HDR thesis.
 }
    \label{fig12}
 \end{figure}

  \begin{figure}
   \centerline{\includegraphics[width=0.98\textwidth,clip=]{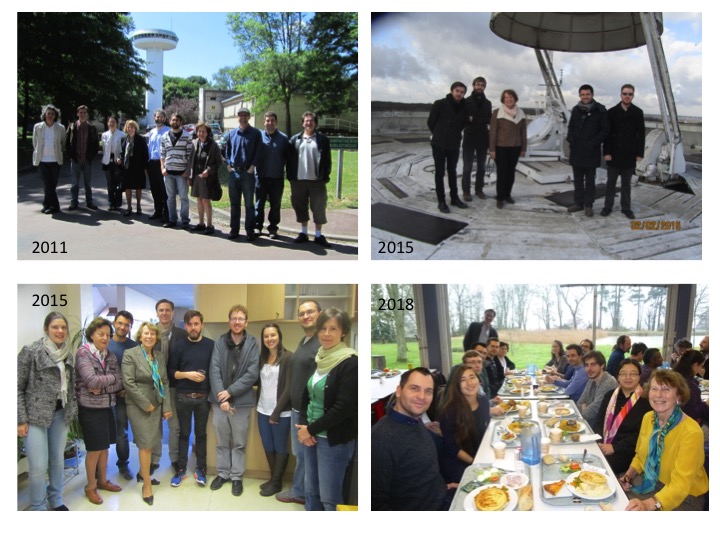}}
 \caption{In front of the solar tower of Meudon, the solar group   in  September 2016:  Gehardo Valori, Guillaume Aulanier,  Rositsa Miteva,  BS, Pascal D\'emoulin, Kevin Dalmasse, Monique Pick, Marc Linton, Hamish  Reid, Etienne Pariat,
% Choe Guennou, Sophie Musset, Francesco Zuccarrello,  Nicole Vilmer, Eoin  Carley, Kostas Moriatis, Pascal D\'emoulin, Carolina Malas-Matamoros, Ludwig Klein, BS., 
top right,  top of the solar tower with the coelostat in 2015:  Eoin Carley, Peter Levens, BS, Francesco Zuccarrello, Stuart Gilchrist, bottom left, in Meudon 2015: Sophie Musset, Monique Pick, F.Zuccarello, BS, G.Aulanier, E. Carley, Stuart Gilchrist, Carolina Malas-Matamoros, Etienne Pariat, Sophie Masson, bottom right, 2018 Meudon cafeteria, from left to right, Stano Gunar, Miho Janvier, Krysztof Barczynski, F.Zuccarello, Jaro Dudik, N.Vilmer, G.Aulanier, S. Masson, E.Pariat, -, Luis Linan, Guiping Ruan, BS.
% bottom left , the part of solar group in 2015: 
 %0 : first row Susan Samuel, Anna Lisa Restante, Lidia van Driel, Petr Heinzel, B.Schmieder, second row: Rositsa Miteva,  Mark Linton, G.Aulanier, P.D\'emoulin, E.Pariat, Arek Berlicki, 
 %bottom right, IAU 340 in Jaipur India  in February 2018: Ramesh Chandra, W.Uddin, BS, Sanjay Gosain.
}
    \label{fig24}
 \end{figure}

     Extreme  solar events  causing  strong disturbances on the Earth became a hot topic  because the satellite Kepler registered extreme events on stars  similar to the Sun as discovered  by the Shibata group \cite{Shibata2013}. Could that happen with the Sun?  Guillaume Aulanier used his OHM simulation and the historical frames of the Sun since 1909 kept in  the Meudon archives  to demonstrate that it could not happen with the present Sun   \cite{Aulanier2013,Schmieder2018b}.   
    
    \section{Conclusion}
    
    Extensive progress on the understanding  of our Sun has been made  during the last 50 years. I realized that our group has been  pioneers of many new  ideas in different domains. Let us quote some of them, the heating of the corona not  by  the acoustic  waves existing in the photosphere, the evaporation in solar flares explaining  bright flare  ribbons, the dynamics of  prominences (counter streaming, multi-threads, tornadoes), the filament  eruption mass loading of  coronal mass ejections, Ellerman bombs and the sea serpentine flux tube  as it is crossing the solar surface. Our pioneer  ideas are often rediscovered ten years later by other groups  when the topic becomes fashionable.  My list of publications  in peer reviews reaches 280  papers today.
    
    This has been possible because of my concept of research including observations and theory.  I created a solar MHD group in Meudon with my previous PhD students. Now we are  training many post docs or students from here and  abroad. I am personally invited to  many countries to give lectures and advice for students. My life was always turned to others: what can I bring to him/her, what can I learn from him or her.  It was a continuous exchange and it is why  I am traveling  to so many countries.   
 During all  these  years  many instruments were developed, in space (SMM, Yohkoh, SOHO, TRACE, Hinode, RHESSI, SDO, IRIS) and on the ground, the French and  Italians built a  telescope in Tenerife, called THEMIS with a multi-line magnetograph  (MTR) and the  {\it Multi-Channel Substrative Double Pass} (MSDP) spectrograph.  Every one of these instruments   brought to us new discoveries.  THEMIS (1996-2016) was a very successful  instrument  giving specific results that no other instruments can provide, e.g., measuring the magnetic field in prominences.  
 %Unfortunately the promotors of THEMIS died too early (Jean Rayrole,  Elisabeth Ribes, Jean Arnaud) or retired (Jean Louis Leroy) and  this instrument without its  promotors became an orphan. Most of the  next generation of scientists around THEMIS  disappeared   the field progressively for personal reasons, I believe,  and no group could be formed around the instrument. 
 Now the instrument is  upgraded with  new adaptive optics, let us hope  a new life for it. 
 
 For the future, in the horizon of the 2020's,  we saw already the launch  of  the Parker Solar Probe  to the Sun, we will have soon the departure of Solar Orbiter  going also  towards the Sun, the first light in  the US  DKIST solar telescope, and perhaps   the construction of EST, the European Solar Telescope,  on the ground. I hope that they will bring also important discoveries on the solar wind, its origin and the processes of acceleration of particles. Our solar group should get  involved in  these instruments.
 
 Nowadays international collaborations are even more important than before because  astrophysics is such a  complex topic, mixing many different physical processes. The new system of post-docs leaving  the group after  two or three 
 years increases our potential of  research but is not as efficient as  collaborations with permanent researchers in other countries who can continue to collaborate. 
 % Research positions in Europe are seldom, in China it is still open for good researchers  and many Chinese coming back from Europe or US get a permanent job in China. China is developing very fast in all the domains and particularly in Solar Physics \cite{Fang2018}. 

    My career was fantastic and is still very lively. I met a lot of friends and worked in wonderful conditions.  I was selected in 2010 among 1000 French scientists to be honored by  having projected my photo on the Pantheon during the Science Days in Paris (Figure \ref{fig12} top right panel).
 
    I got the awards of  "Chevalier and Officier des Palmes acad\'emiques"  in 2010  and  the awards of "Chevalier and Officier de la L\'egion d'Honneur'' in 2012 for my involvement in education and research (Figure \ref{fig12} top left panel).
    
     I have educated many students (master and PhD) and post-docs in solar physics. We celebrated successes with champagne each time (Figure \ref{fig12} bottom  panels). I apologize that I could not quote all of them.  I want to thank all my colleagues from the past and those of today. The ambience is still very friendly and  I  try to communicate my enthusiasm of the research to my younger colleagues
    from all over the world (Figure \ref{fig24}). During my thesis  research, I have to  define the  problems to resolve. It was  a white page on my desk  every day and I have to think during all the day what to compute and write. Now I have so many questions  in solar physics to answer that I have not enough  time and people   around me to achieve   the task.

     I have many other  scientific  activities  as referee of many scientific papers and many proposals (EU, Australian, Belgium FWO). I am member of the SAF committee (Soci\'et\'e Astronomique de France) and animated public events such as    eclipses,  and the passage of Mercury in front of the Sun. I  also like dancing, music, listening operas, nature, I like to go to seashore
when it is summer time, play the piano, grow flowers and enjoy their beauty.

    The salary of astronomer is not so high and it was not easy  to travel with kids at home but
    I thank my family for  accepting  my life as it was. 
      %    Coming back to the question that I had to resolve for my thesis: are the acoustic waves responsible of the heating of the corona to one million degree, I realized that I worked on that question during all my professional life.  We have shown that the corona is highly magnetized and full of electric field. The sun is a big magnet with a north pole and and south pole but also we see at the surface many smaller magnets of different sizes like the active region with two sunspot one with positive and one with negative polarities and very tiny bipoles in the quiet sun. The bipoles in the photosphere are the footprints of magnetic field lines forming loops.
 %   Vector magnetographs like THEMIS or in space HMI provide the values of the magnetic field. In the corona there is no real way to measure it. We need to extrapolate the magnetic field of to make simulations.   It is what we did in all our papers. We note that any photospheric motions (due to convection) stress the files and the  flux rope are twisted.  This stress or twist lead to generation of magnetic waves or to reconnection. In both cases  the plasma is heated, particles are expelled and accelerated  and evacuated to the heliosphere. During the last decade we understood that beyond the corona there is a tiny plasma of weak density called the solar wind. In the solar wind accelerated particles and  coronal mass ejection are traveling in the solar wind.  What is the effects of all these ionized and magnetized plasma arriving in the Earth environment. That is the new question of the XXI century to resolve?
      I have  fantastic grandchildren that many of my colleagues know because they accompany  me in   meetings in  different countries (Camille, Adrien, Gr\'egoire, Gautier, Capucine, Baptiste). They are interested  in watching the  solar eclipses,  observing Jupiter and Saturn with a large telescope. The two   youngest grandchildren (Cl\'ementine and Margot) hope that it will be soon  their turn.

%% Table
%
% \begin{table}
% \caption{}%\label{tbl:?}
% \begin{tabular}{}     
% \hline
% \multicolumn{2}{c}{<>}
% <data>
% \hline
% \end{tabular}
% \end{table}

%%%%%%%%%%%%%%%%%%%%%%%%%%%%%%%%%%%%%%%%%%%%%%%%%%%%%%%%%%%%%%%%%%%%%%%%%%%
%% Appendix
%
% \appendix   {Figure 1: Flare Genesis Experiment results : (a)  magnetic field in the photosphere,  (b) the currents in the photosphere,  (c) Ellerman bombs,  Jointed observations (d) intensity of the transition region with TRACE (d) in the corona ,  (e)  very hot loops in the corona.With Yohoh/SXT (Pariat et al 2004, Schmieder et al 2004).}

%%%%%%%%%%%%%%%%%%%%%%%%%%%%%%%%%%%%%%%%%%%%%%%%%%%%%%%%%%%%%%%%%%%%%%%%%%%
%% Acknowledgements
%
% \begin{acks}
%
% \end{acks}

%%% %%%%%%%%%%%%%%%%%%%%%%%%%%%%%%%%%%%%%%%%%%%%%%%%%%%%%%%%%%%
%% Bibliography
%
% Using BibTeX
%
\mbox{}~\\
% \bibliographystyle{spr-mp-sola}
% %\bibliographystyle{spr-mp-sola-cnd} %% Alternative style: no title, no concluding page
% \bibliography{bibliography1}  

 \IfFileExists{\jobname.bbl}{} {\typeout{}
\typeout{***************************************************************}
\typeout{***************************************************************}
\typeout{** Please run "bibtex \jobname" to obtain the bibliography}
\typeout{** and re-run "latex \jobname" twice to fix references}
\typeout{***************************************************************}
\typeout{***************************************************************}
\typeout{}}
%
% Without BibTeX 
% \begin{thebibliography}{}
% \bibitem[\protect\citeauthoryear{Author}{Year}]{key}
%   <bibliographical entry>
%
% \bibitem[\protect\citeauthoryear{}{}]{}
%   
%  
% \end{thebibliography}

\end{article} 
\end{document}